\documentclass[a4paper]{article}

\usepackage{amssymb}   
\usepackage{amsmath}   
\usepackage{graphicx} 
\usepackage{psfrag}   
\usepackage{pstricks}  
\usepackage{multido}   
\usepackage{tikz}      
\usetikzlibrary{arrows,decorations.pathmorphing,%
  backgrounds,positioning,fit}
\usepackage{pgfmath}   
\usepackage[colorlinks]{hyperref}

\title{From entanglement renormalisation to the disentanglement of
  quantum double models}

\date{\today}

\author{%
  Miguel Aguado\footnote{miguel.aguado@mpq.mpg.de} \\
  Max-Planck-Institut f\"ur Quantenoptik \\
  Hans-Kopfermann-Str.~1. D-85748 Garching, Germany, \\
  and The Kavli Institute for Theoretical Physics, \\
  University of California, Santa Barbara CA 93106-4030, USA.}

\begin{document}

\maketitle


\begin{abstract}
  We describe how the entanglement renormalisation approach to
  topological lattice systems leads to a general procedure for
  treating the whole spectrum of these models, in which the
  Hamiltonian is gradually simplified along a parallel simplification
  of the connectivity of the lattice.  We consider the case of
  Kitaev's quantum double models, both Abelian and non-Abelian, and we
  obtain a rederivation of the known map of the toric code to two
  Ising chains; we pay particular attention to the non-Abelian models
  and discuss their space of states on the torus. Ultimately, the
  construction is universal for such models and its essential feature,
  the lattice simplification, may point towards a renormalisation of
  the metric in continuum theories.
\end{abstract}

\begin{flushleft}
  NSF-KITP-10-163
\end{flushleft}

\begin{flushleft}
\end{flushleft}

\clearpage

\tableofcontents

\clearpage

\section{\label{sec:intro}%
  Introduction}

The application of tensor network methods has provided deep insight
into topologically ordered systems.  Tensor networks provide a setting
where the exotic characteristics of topological order (topology
dependent ground level degeneracy, local indistinguishability of
ground states, topological entanglement entropy, interplay with
renormalisation, unusual realisation of symmetries) can be exactly
studied.  On the other hand, tensor network methods are sufficiently
flexible for studying the excitations of these systems; in 2D, as is
known, the excitations are quasiparticles with anyonic exchange
statistics, which is the basis of Kitaev's topological quantum
computer architecture \cite{Kitaev:1997}.

Among lattice models with topological order, Kitaev's quantum double
models \cite{Kitaev:1997} have a distinguished position.  They were
the first models in which the possibilities of the topological setting
for quantum computation purposes were discussed, and include anyon
models universal for quantum computation by braiding; on the other
hand, quantum double models bear an intimate relation to discrete
gauge theories, exhibiting a rich group-theoretical and algebraic
structure that pervades the study of their ground level and
excitations, and in particular determines the properties of the anyons
and their computational power.  While they are not as general as
string-net models \cite{stringnet}, the models that describe all
doubled topological fixed points on the lattice, steps towards a
reformulation of the latter in the spirit of quantum double models
have been taken in \cite{BA, BMCA, BCKA}.  Recent progress in the
perturbative production of quantum double codes via two-body
interactions has been reported in \cite{brell}.

Tensor network methods were first applied to topological models on the
lattice in \cite{power}, where an exact projected entangled-pair state
(PEPS) representation of a ground state of the toric code was given.
In recent years, a growing number of tensor network representations
for topological states have been developed following the lead of
\cite{power} (e.g. \cite{AguadoVidal, Koenig, BAV, GuTopo, BMCA}),
which are notably exact for ground states of fixed-point Hamiltonians
(quantum doubles, string-nets), and moreover lend themselves to deep
analysis of their symmetries \cite{Schuch}.  Tensor network methods
for systems with anyonic quasiparticles, independent of any
microscopic structure, have also been developed \cite{KoenigAnyonMERA,
  Pfeifer}.

The multi-scale entanglement renormalisation ansatz (MERA) is a
particular kind of tensor network structure, developed by G.~Vidal
\cite{meraa, merab}, which builds on the coarse-graining procedure
typical for real-space renormalisation flows, but introduces layers of
unitary operators between coarse-graining steps so as to reduce
interblock entanglement, a procedure called entanglement
renormalisation (ER).  In practice, ER led to tractable MERA
representations and algorithms for ground states of critical systems
in 1D, traditionally hard for tensor network methods, and the
numerical applications of the method now encompass a wide class of
lattice and condensed-matter models both in 1D and in higher
dimensions.  ER was first applied to topological lattice models in
\cite{AguadoVidal}, where exact MERAs were given for ground states of
quantum double models, both Abelian and non-Abelian.  In \cite{Koenig}
ground states of string-net models were also written in MERA form.

While understanding ground states of many-body quantum systems is
certainly of the utmost importance, it is a bonus for a method to be
able to explore at least the low-lying excitation spectrum.  The
purpose of this paper is to show how the principles of the ER approach
to quantum double models can be extended to account for the whole
spectrum in an exact way.  The picture that emerges is somehow
complementary to entanglement renormalisation: by giving up the
coarse-graining, that is, using unitary tensors throughout, the ER
method turns into a reorganisation of degrees of freedom which is
essentially graph-theoretical and proceeds by modifying the lattice by
keeping the topological Hamiltonian the same at each step; this
procedure yields finally a Hamiltonian consisting essentially of
one-body terms, or an essentially classical one-dimensional spin chain
model.

Naturally, the structure of the tensor network with unitary operations
defines a unitary quantum circuit, which in this case simplifies the
structure of the Hamiltonian down to mostly one-body terms.  In this
sense, the construction can be regarded as an instance of the broad
idea of quantum circuits diagonalising Hamiltonians, introduced in
\cite{Latorre:circuit}.

In this vein, we obtain an explicit (and geometrically appealing)
construction of the well known mapping of the toric code to two
classical Ising chains; but we also cover the non-Abelian cases.
Remarkably, the construction is virtually universal for all quantum
double models, and can be understood as a series of moves simplifying
the structure of the lattice into a series of `bubbles' and `spikes',
each one of which talks only to one qudit; each quantum double model
possesses a canonical set of operations transforming the model on one
lattice into the same model on the next lattice.  Since the lattice
structure can be considered as the discretisation of a metric, this
might open the door to speculations about a continuum counterpart
(including, perhaps, dimensional reduction in going from the
topological Hamiltonian in 2D to one-dimensional classical chains).
We remark that the recent paper \cite{Luca} uses related ideas to find
numerical methods applicable to lattice gauge theories.

The structure of the paper is as follows.  The Abelian case of the
toric code is discussed in section \ref{sec:toric}, where we recall
the basic elements of the construction of \cite{AguadoVidal}, and
proceed to a detailed presentation of the disentangling method.  In
section \ref{sec:qdoubles}, the analogous construction is developed
for general quantum double models; particular attention is given to
the models on the torus, where the structure of the Hilbert space has
to be taken into account carefully.  Section \ref{sec:discussion}
contains a discussion and conclusions.  In appendix
\ref{sec:appendix:dgmodels} some algebraic background underlying
quantum double models is provided, and appendix
\ref{sec:appendix:dsthree} concentrates on the simplest non-Abelian
quantum double, that of the group $\mathrm{S}_3$.

\clearpage

\section{\label{sec:toric}%
  Entanglement renormalisation of the toric code}

We begin by describing the construction in the simplest Abelian case,
namely the toric code.  This will allow us to make a connection with
known results, namely the mapping to a pair of classical Ising chains.

\subsection{\label{subsec:toric:elementary}%
  Elementary moves for the toric code}

We start from an arbitrary two-dimensional lattice $\Lambda$, with
qubits sitting at the bonds.  We do not specify the topology yet.  The
Hamiltonian, to begin with, has the familiar form of Kitaev's toric
code \cite{Kitaev:1997}:
\begin{equation}\label{disent:hlambda}
  H_\Lambda
=
 - \, \sum_P B_P
 - \, \sum_V A_V
  \; ,
\end{equation}
with
\begin{equation}
  B_P
=
  \bigotimes_{ i \in P } Z_i \; ,
\quad
  A_V 
=
  \bigotimes_{ j \in V } X_j
\end{equation}
mutually commuting stabilisers with support on plaquettes $P$ and
vertices $V$.  Any code state $\lvert \psi \rangle$ (i.e., any ground
state of $H_\Lambda$) satisfies the local conditions $B_P \lvert \psi
\rangle = \lvert \psi \rangle$, $A_V \lvert \psi \rangle = \lvert \psi
\rangle$ for all $P$, $V$; whatever information it carries is purely
topological, encoded in the eigenvalues of nonlocal operators.

The entanglement renormalisation scheme proposed in \cite{AguadoVidal}
uses two kinds of elementary operations introduced in \cite{Dennis},
the plaquette move and the vertex move.  In that work only code states
were considered, and the result of applying elementary moves was to
decouple qubits from the code in known states.  More precisely,
plaquette moves (P-moves) decouple a qubit $q$ from the lattice by
putting it in a product state $\lvert 0 \rangle$ with the rest of the
system:
\begin{equation}\label{bubbles:plaquettemovestates}
  \lvert \text{code state; $w$} \rangle_\Lambda
\stackrel{P}{\mapsto}
  \lvert \text{code state; $w$} \rangle_{\Lambda'}
 \otimes
  \lvert 0 \rangle_q
  \; ,
\end{equation}
where $\Lambda'$ is the lattice resulting from the deletion of the
edge containing $q$, therefore fusing the two plaquettes to which $q$
belonged into one plaquette.  Here $w$ stands for any nonlocal
information in the code state, which is preserved provided the
plaquette move is purely a bulk operation, that is, it acts on a
contractible region.  Vertex moves (V-moves) effect a similar
simplification,
\begin{equation}\label{bubbles:vertexmovestates}
  \lvert \text{code state; $w$} \rangle_\Lambda
\stackrel{V}{\mapsto}
  \lvert \text{code state; $w$} \rangle_{\Lambda''}
 \otimes
  \lvert + \rangle_q
  \; ,
\end{equation}
with $\Lambda''$ obtained by contracting the edge containing $q$ into
a vertex (subsuming the two vertices of the contracted edge).  In both
cases the topological information carried by the code state is
unchanged, if the moves have support on a contractible region.

The programme of entanglement renormalisation can be carried out
explicitly and exactly for the code states, as shown in
\cite{AguadoVidal}, using P-moves and V-moves organised into
disentanglers and isometries (in the latter case the decoupled qubits
being discarded from the system).  This gives rise to an exact MERA
ansatz for the code states, whereby the top of the tensor network
contains only topological degrees of freedom; for instance, in the
case on the torus, these topological degrees of freedom are two qubits
on edges along noncontractible loops of $\mathrm{T}^2$.

\begin{figure}[htbp]
\centering
\includegraphics{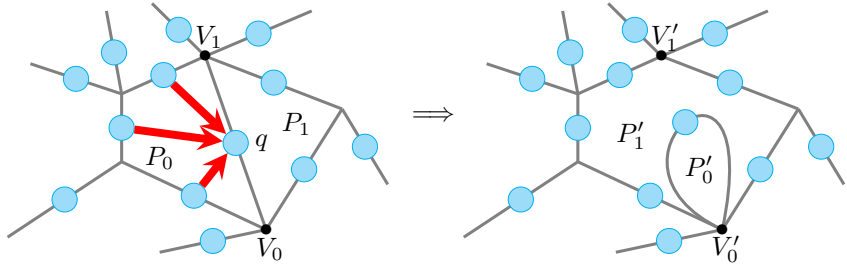}
\caption{\label{bubbles:fig:recoupleplaq} {\small \textbf{Deformation
      of the lattice after a P-move}.  The P-move consists of CNOTs
    (red arrows).  Qubit $q$ ``decouples'' if in the ground state, but
    the complete picture is that it remains in the deformed lattice,
    with its own plaquette (a bubble containing the flux initially in
    the left subplaquette).  To be consistent with the non-Abelian
    generalisation to be discussed later on, the bubble is attached to
    one of the vertices talking to $q$.}}
\end{figure}

Let us consider more in detail a P-move $U_{\mathrm{plaq}}$ (figure
\ref{bubbles:fig:recoupleplaq}, left).
\begin{figure}[htbp]
\centering
\includegraphics{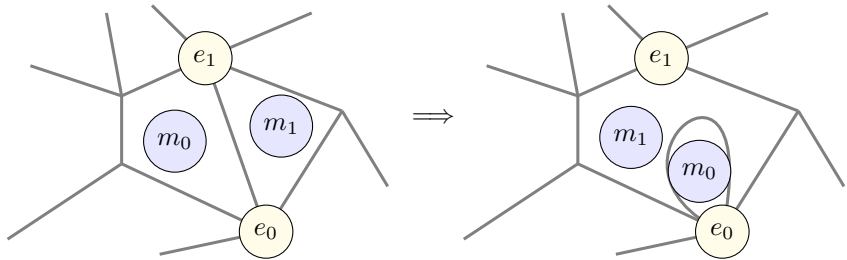}
\caption{\label{bubbles:fig:recoupleplaqflux} {\small \textbf{Charges
      under a P-move.} Redistribution of electric charges $e_i$ and
    magnetic fluxes $m_i$ in the P-move of figure
    \ref{bubbles:fig:recoupleplaq}.  Note that the redistribution
    would be different (swapping fluxes $m_0$, $m_1$ in the deformed
    lattice) if the control qubits there belonged to the plaquette
    $P_1$, although for the ground states the effect is the same.
    Each deformed vertex contains the same electric charge as its
    undeformed predecessor.}}
\end{figure}
This consists of simultaneous CNOTs with target qubit $q$ and control
qubits in the remaining edges around one of the two plaquettes $P_0$,
$P_1$ containing $q$ (say, $P_0$).  Call $P$ the plaquette obtained
from $P_0$ and $P_1$ by deleting the bond for $q$.  Only four terms in
the Hamiltonian change:
\begin{equation}\label{bubbles:hchange}
  U_{\mathrm{plaq}}
  (
    B_{ P_0 } + B_{ P_1 }
   +
    A_{ V_0 } + A_{ V_1 }
  )
  U_{\mathrm{plaq}}^\dagger
=
  Z_q + Z_q B_P
 +
  A_{ V'_0 } + A_{ V'_1 }
  \; ,
\end{equation}
where vertices $V_0$ and $V_1$ talk to qubit $q$, and $V'_0$, $V'_1$
are the new vertices after deletion of the bond containing $q$.

But this calls for a more general interpretation than just the
coarse-graining (\ref{bubbles:plaquettemovestates}) of the ground
level.  The first term in the rhs of (\ref{bubbles:hchange}) describes
qubit $q$ having eaten the flux in plaquette $P_0$, but not otherwise
subject to vertex constraints; the second term contains all qubits in
$P_0$ and $P_1$, including $q$ counted once, so the new Hamiltonian is
\emph{not} a toric code with $q$ removed, but a toric code with a
deformation of the lattice still containing $q$ \emph{in a bubble},
i.e., in its own plaquette $P'_0$ contributing to the total flux
(figure \ref{bubbles:fig:recoupleplaq}, right).  Note that we draw
$P'_0$ attached to one of the vertices of the bond previously
containing $q$; this is needed for a consistent generalisation to
non-Abelian cases.  At any rate, $q$ is not subject to any vertex
constraint since it talks \emph{twice} to this vertex $V'_0$, and $X^2
= 1$.

The effect of the P-move in terms of charges and fluxes is drawn in
figure \ref{bubbles:fig:recoupleplaqflux}.  The electric charges
remain at their vertices, while the bubble $P'_0$ of $q$ contains the
flux across $P_0$, and the complementary new plaquette $P'_1$ contains
the flux previously in $P_1$.  Note that while the effect on the
ground level is symmetric with respect to the choice of control
qubits, this is not the case when there are nontrivial fluxes.

Note also that the total charge and flux of the entire region affected
by the local operation $U_{\mathrm{plaq}}$ are conserved, as they
should be.

Elementary V-moves are entirely dual to P-moves in the toric code
(and, more generally, in any Abelian quantum double model).  To be
precise, the duality is established by going over to the dual lattice
and performing a global Hadamard, thus swapping $X$ and $Z$ operators.

\begin{figure}[htbp]
\centering
\includegraphics{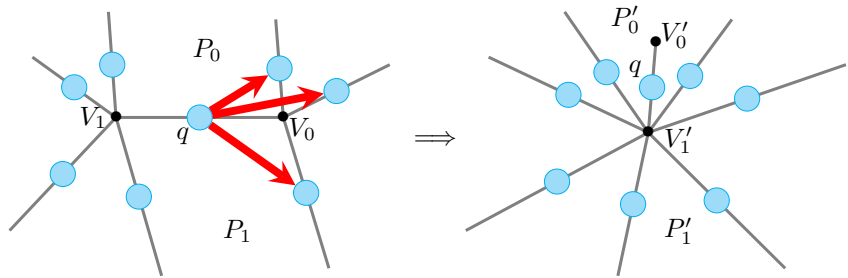}
\caption{\label{bubbles:fig:recouplevert} {\small \textbf{Deformation
      of the lattice after a V-move.} As before, red arrows stand for
    CNOTs.  Qubit $q$ belongs, after the operation, to a spike (a bond
    with one of its vertices not connected to any other bond).}}
\end{figure}

\begin{figure}[htbp]
\centering
\includegraphics{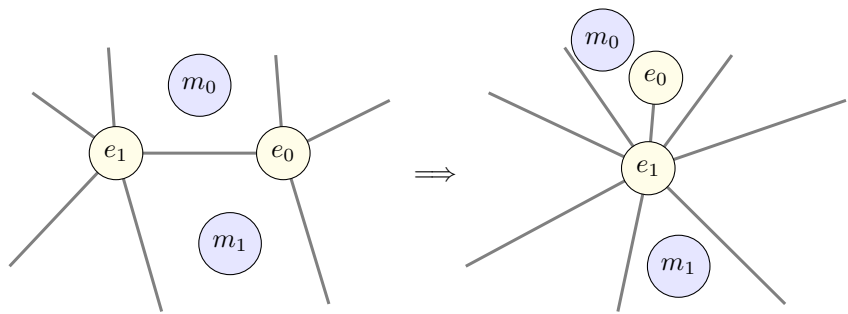}
\caption{\label{bubbles:fig:recouplevertflux} {\small \textbf{Charges
      under a V-move.} Redistribution of charges and fluxes in a
    V-move.  Again, note that this is not symmetric in the vertices.
    Had the CNOTs in figure \ref{bubbles:fig:recouplevert} acted on
    the qubits surrounding $V_1$, the electric charges $e_0$, $e_1$ in
    the deformed lattice would have been swapped.}}
\end{figure}

Figure \ref{bubbles:fig:recouplevert} shows the effect of such a move,
$U_\mathrm{vert}$.  The operation consists of CNOTs with the control
qubit $q$ in the bond from vertex $V_0$ to $V_1$ (the one to
`decouple' in state $\lvert + \rangle$ if in a ground state) and
targets those sharing with $q$ one of the vertices, say $V_0$.  The
result is that the Hamiltonian is mapped to the Hamiltonian of a
lattice where the bond of $q$ has been contracted, identifying its two
vertices into a new vertex $V'_1$, to which is attached a `spike',
i.e., a bond, containing $q$, whose other vertex $V'_0$ is not
connected to any other bonds.  As shown in figure
\ref{bubbles:fig:recouplevertflux}, the fluxes in the new plaquettes
$P'_0$ and $P'_1$ are inherited from those in $P_0$ and $P_1$, while
the electric charge at vertex $V'_0$ contains the charge previously at
vertex $V_0$, and the vertex $V'_1$ at the end of the spike contains
the charge of $V_1$.
%

\subsection{\label{subsec:toric:example}%
  Disentangling the Hamiltonian: a simple example}

With the set of elementary moves at hand, we can disentangle the
Hamiltonian by simplifying the structure of the lattice.

Think of a toric lattice, to be specific an $L_x \times L_y$ square
lattice $\Lambda_0$ with periodic boundary conditions, on which a
toric code $\mathrm{\mathbf{TC}} ( \Lambda_0 )$ is defined.  By using
P-moves and V-moves as described in the previous section, the lattice
gets deformed into a sequence of lattices $\Lambda_n$, $n = 1, \, 2,
\, \ldots$, with the same number of vertices, edges, and plaquettes,
and hence the same \emph{topology}, but increasingly simpler
connectivity, and keeping the lattice system a toric code
$\mathrm{\mathbf{TC}} ( \Lambda_n )$ at each step.  States flow in
this process just by redistributing the charges and fluxes.

The lattice can be simplified to reach a particularly appealing form,
to which corresponds a Hamiltonian consisting of three sectors:
\begin{itemize}
\item %
  A \emph{trivial} sector corresponding to two qubits encoding the
  topological degrees of freedom.
\item %
  An Ising chain of $L_x L_y - 1$ qubits encoding the structure of
  magnetic (plaquette) fluxes.
\item %
  An Ising chain of $L_x L_y - 1$ qubits encoding the structure of
  electric (vertex) charges.
\end{itemize}

To understand this, note that the minimal toric square lattice
consists of two links along two nonequivalent nontrivial loops on
$\mathrm{T}^2$, beginning and ending at a unique vertex, and defining
a unique plaquette.  In this minimal case, the Kitaev Hamiltonian is
trivial and all computational basis states are Kitaev logical states,
there being no room for excitations in the Hilbert space.

On performing P-moves and V-moves, the topology of the lattice is left
unchanged, but the number of nontrivial cycles along links in the
lattice can be reduced to the minimal case, whereby two qubits carry
all topological information.  Information about excitations in the
original lattice will appear in the rest of the code qubits.

\begin{figure}[htbp]
\centering
\includegraphics{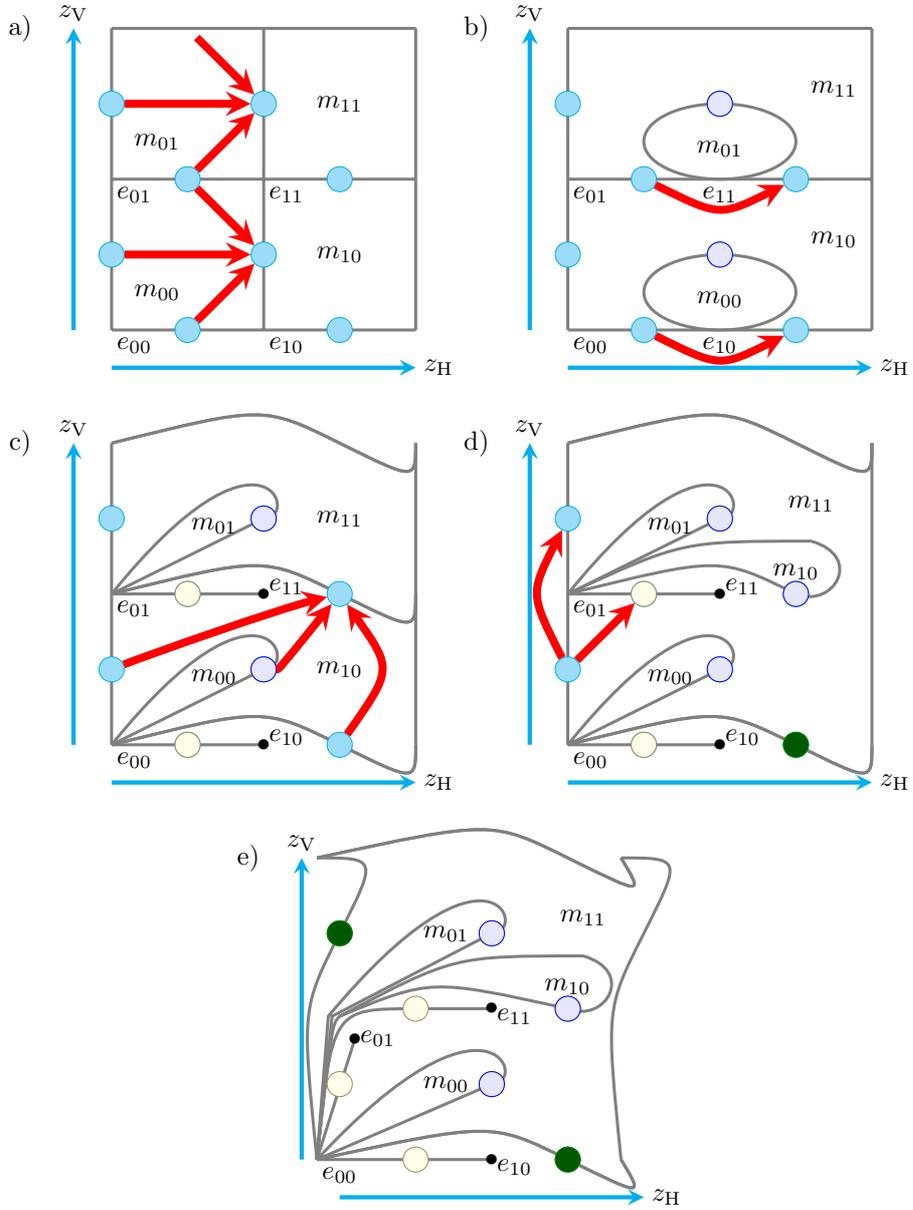}
\caption{\label{fig:twobytwotorus} {\small \textbf{Simplification of
      the lattice for a $2 \times 2$ toric code.}  The end lattice
    features two topologically nontrivial links and a series of
    bubbles and spikes, all attached to the same vertex.  We have kept
    qubits in their places and only deformed the lattice links.  Red
    arrows stand for CNOTs and point from control to target qubits.
    Dark qubits are those that hold a full qubit of topological
    information.  Other colours are used for qubits that hold the flux
    (medium dark) or charge (light) of a plaquette or vertex of the
    original lattice.}}
\end{figure}

\begin{figure}[htbp]
\centering
\includegraphics{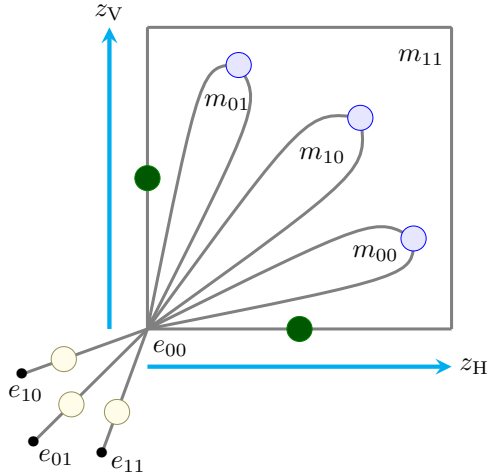}
\caption{\label{fig:twobytwocleaner} {\small \textbf{A clearer picture
      of the disentangled $2 \times 2$ toric code.}  Light qubits
    carry electric information, medium dark qubits carry magnetic
    information, and dark qubits carry topological information.  The
    Hamiltonian for this code is
    (\ref{bubbles:twobytwohamiltonian}).}}
\end{figure}

Let us illustrate the simplification in the case of a toric lattice
with four plaquettes and four vertices.  The steps are depicted in
figure \ref{fig:twobytwotorus}.
We consider a basis state of the Hilbert space determined by magnetic
fluxes $m_{ij}$ for plaquettes $\mathrm{p}_{ij}$ and electric charges
$e_{ij}$ at vertices $\mathrm{v}_{ij}$, taking values in $\{ \pm 1 \}$
(being the eigenvalues of the corresponding plaquette or vertex
operator).  Indices $i, \, j$ run from 1 to 2 labelling the horizontal
and vertical logical coordinates of the plaquettes and vertices.  By
overall charge neutrality, $\prod_{ij} m_{ij} = +1 = \prod_{ij}
e_{ij}$.  Topological information is encoded in two nonequivalent
nontrivial loops, the eigenvalues of the corresponding logical $Z$
operators being $z_{\mathrm{H}}$, $z_{\mathrm{V}}$.  After a series of
P-moves and V-moves, the topological data are carried by the two
qubits shown in dark in figure \ref{fig:twobytwotorus}, which will be
denoted as $q_{V, \, H}$.  Qubits carrying magnetic information will
be referred to as $q_{M, \, i}$, $i = 1, \, 2, \, 3$, and are those
whose links enclose a solitary plaquette.  Qubits carrying electric
information are shown in light colour in figure
\ref{fig:twobytwotorus}, will be denoted as $q_{E, \, i}$, $i = 1, \,
2, \, 3$, and are those whose link has a free vertex.  The final
Hamiltonian is
\begin{align}\label{bubbles:twobytwohamiltonian}
\nonumber
  H_{\text{final}}
&=
 - \,
  \sum_{ i = 1 }^3
  Z ( q_{M, \, i} )
 -
  Z ( q_{M, \, 1} ) Z ( q_{M, \, 2} ) Z ( q_{M, \, 3} )
 \\
&
\quad
 - \,
  \sum_{ i = 1 }^3
  X ( q_{E, \, i} )
 -
  X ( q_{E, \, 1} ) X ( q_{E, \, 2} ) X ( q_{E, \, 3} )
  \; .
\end{align}
The precise redistribution of fluxes and charges will be discussed in
general later, but for the moment let us remark that, as promised,
there are no terms in $H_{\text{final}}$ associated with $q_{V, \,
  H}$, and that the magnetic sector (terms affecting the $q_{M, \,
  i}$, i.e., involving $Z$s) and the electric sector (terms affecting
the $q_{E, \, i})$, i.e., involving $X$s) are each intimately related
to an Ising chain Hamiltonian.  A cleaner illustration of the lattice
structure associated with Hamiltonian
(\ref{bubbles:twobytwohamiltonian}) is given in figure
\ref{fig:twobytwocleaner} (the Abelian nature of the code allows for
generous rearrangement of bubbles and spikes).

\begin{figure}[htbp]
\centering
\includegraphics{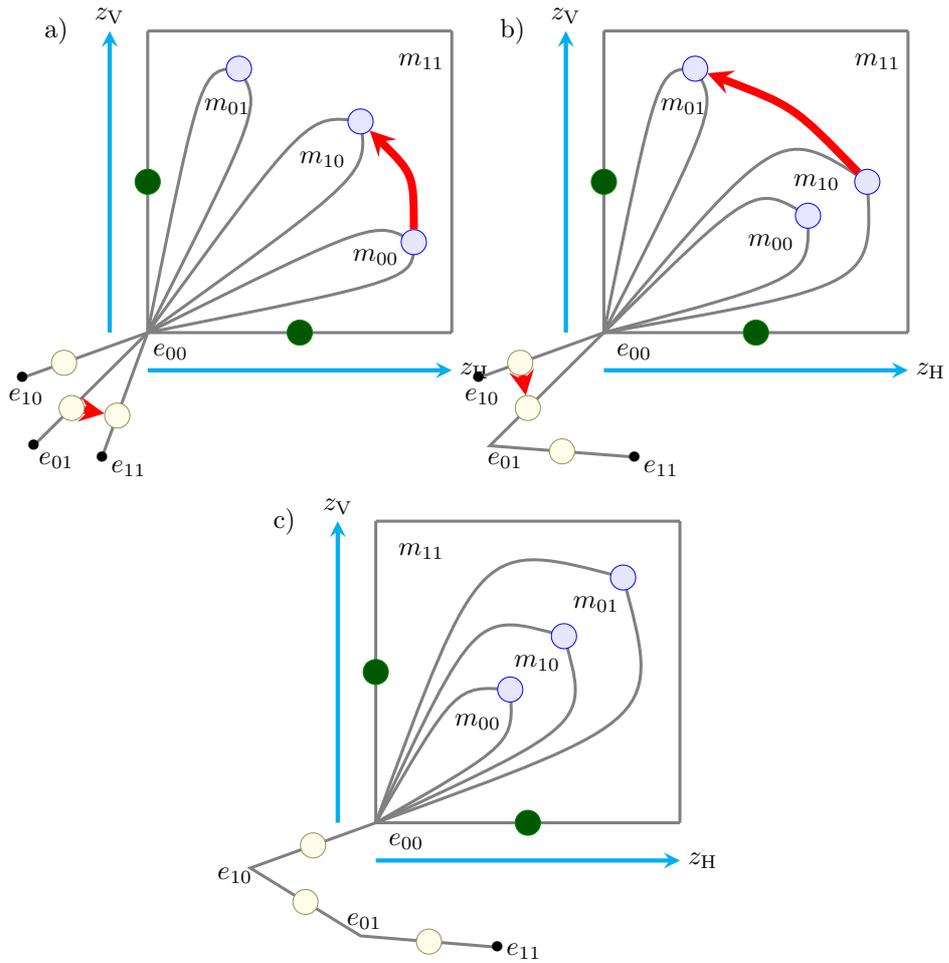}
\caption{\label{fig:twobytwotoising} {\small \textbf{From the
      disentangled toric code to Ising chains.}  The sequence of steps
    to bring the lattice explicitly to the form where the toric code
    Hamiltonian is the sum (\ref{bubbles:twobytwoisinghamiltonian}) of
    two noninteracting Ising chains.  Red arrows, as usual, stand for
    CNOTs.}}
\end{figure}

The correspondence with the Ising chains can be made explicit by
especially simple P-moves and V-moves acting on the configuration of
figure \ref{fig:twobytwocleaner} by direct application of stabiliser
formalism results.  These are shown in figure
\ref{fig:twobytwotoising}.
Denote as $\mathcal{C} ( t \leftarrow c )$ the CNOT operation with
control $c$ and target $t$.  Starting from
(\ref{bubbles:twobytwohamiltonian}), the corresponding operation is
\begin{equation}\label{bubbles:operationtoising}
  \mathcal{C} ( q_{E, 2} \leftarrow q_{E, 1} )
  \mathcal{C} ( q_{E, 3} \leftarrow q_{E, 2} )
  \mathcal{C} ( q_{M, 3} \leftarrow q_{M, 2} )
  \mathcal{C} ( q_{M, 2} \leftarrow q_{M, 1} )
  \; ,
\end{equation}
whose application yields the doubled Ising Hamiltonian:
\begin{align}\label{bubbles:twobytwoisinghamiltonian}
\nonumber
  H_{\text{final}}
&=
 - \,
  Z ( q_{M, \, 1} )
 - \,
  Z ( q_{M, \, 1} ) Z ( q_{M, \, 2} ) 
 - \,
  Z ( q_{M, \, 2} ) Z ( q_{M, \, 3} ) 
 - \,
  Z ( q_{M, \, 3} ) 
\\
&
\quad
 - \,
  X ( q_{E, \, 1} )
 - \,
  X ( q_{E, \, 1} ) X ( q_{E, \, 2} ) 
 - \,
  X ( q_{E, \, 2} ) X ( q_{E, \, 3} ) 
 - \,
  X ( q_{E, \, 3} ) 
  \; .
\end{align}
%

\subsection{\label{subsec:toric:fullstructure}%
  Disentangling the Hamiltonian: full structure}

\begin{figure}[htbp]
\centering
\includegraphics{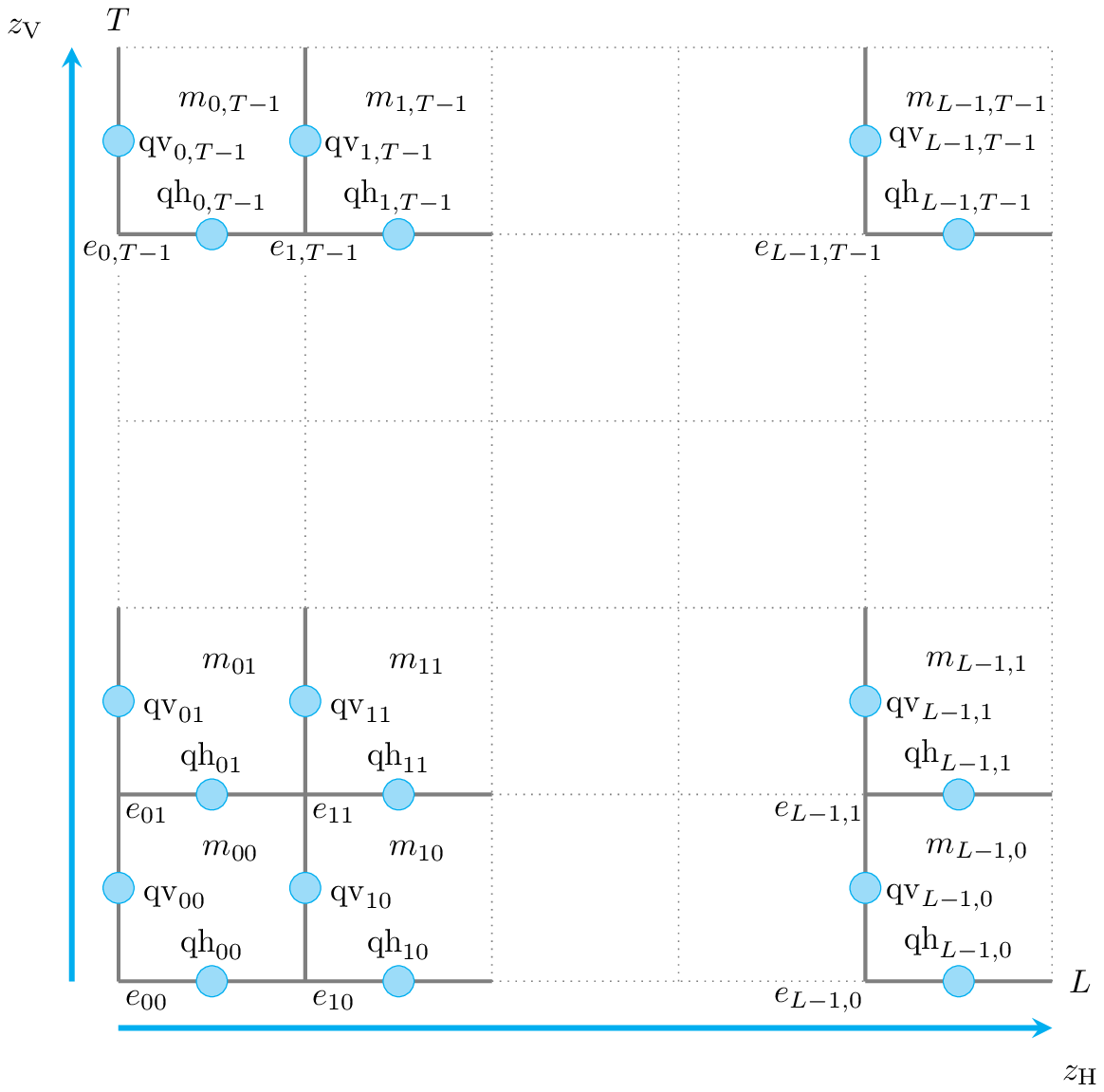}
\caption{\label{fig:fullstructurefull} {\small \textbf{General toric
      code simplification: setting.}  The starting point for the
    discussion of the full structure of the procedure of
    simplification of the toric code.  Here qubits are split into two
    groups according to whether they belong to a horizontal bond
    ($\mathrm{qh}_{i, j}$) or a vertical bond ($\mathrm{qv}_{i, j}$)
    in the original $L \times T$ toric lattice.  Electric charges
    $e_{ij} = \pm 1$ at the vertices and magnetic fluxes $m_{ij} = \pm
    1$ across the plaquettes are labelled as well by obvious row and
    column indices, $i = 0, \, 1, \, \ldots, \, L-1$ and $j = 0, \, 1,
    \, \ldots, \, T-1$.  Overall charge neutrality on the torus means
    $\prod_{ij} m_{ij} = 1 = \prod_{ij} e_{ij}$.}}
\end{figure}

As for the explicit moves and redistribution of charges and fluxes,
necessary to understand the unravelling of the Hamiltonian in general,
consider an $L \times T$ toric square lattice as shown in figure
\ref{fig:fullstructurefull}.
We label vertices according to their coordinates as $\mathrm{v}_{ij}$,
where $i$ runs from $0$ to $L-1$ and $j$ runs from $0$ to $T-1$.
Plaquette $\mathrm{p}_{ij}$ has vertex $\mathrm{v}_{ij}$ as its lower
left corner.  Qubits in horizontal links are denoted
$\mathrm{qh}_{ij}$, with vertex $\mathrm{v}_{ij}$ to their left, and
qubits in vertical links as $\mathrm{qv}_{ij}$, pointing upwards from
vertex $\mathrm{v}_{ij}$.

Consider the basis of simultaneous eigenstates of the set of all
plaquette operators (eigenvalues $m_{ij}$ with $\prod_{ij} m_{ij} =
1$) and vertex operators (eigenvalues $e_{ij}$ with $\prod_{ij} e_{ij}
= 1$), together with logical $Z$ operators $Z_H = \prod_{i=0}^{L-1}
Z_{\mathrm{qh}_{i,0}}$ (eigenvalue $z_H$) and $Z_V = \prod_{j=0}^{T-1}
Z_{\mathrm{qv}_{0, j}}$ (eigenvalue $z_V$).

The process of simplification of the lattice will be described by the
P-moves and V-moves applied to a basis state to bring the lattice to a
form generalising figure \ref{fig:twobytwocleaner}, and the
identification of the qubits carrying the eigenvalue of each vertex,
plaquette, and logical original operator at that stage.

There is a huge amount of freedom in the choice of the process.  I
will be describing one possible choice, not optimal in any sense, but
allowing a systematic treatment.

\paragraph{\label{para:rows}%
  Step 1: Create bubbles along rows}

\begin{figure}[htbp]
\centering
\includegraphics{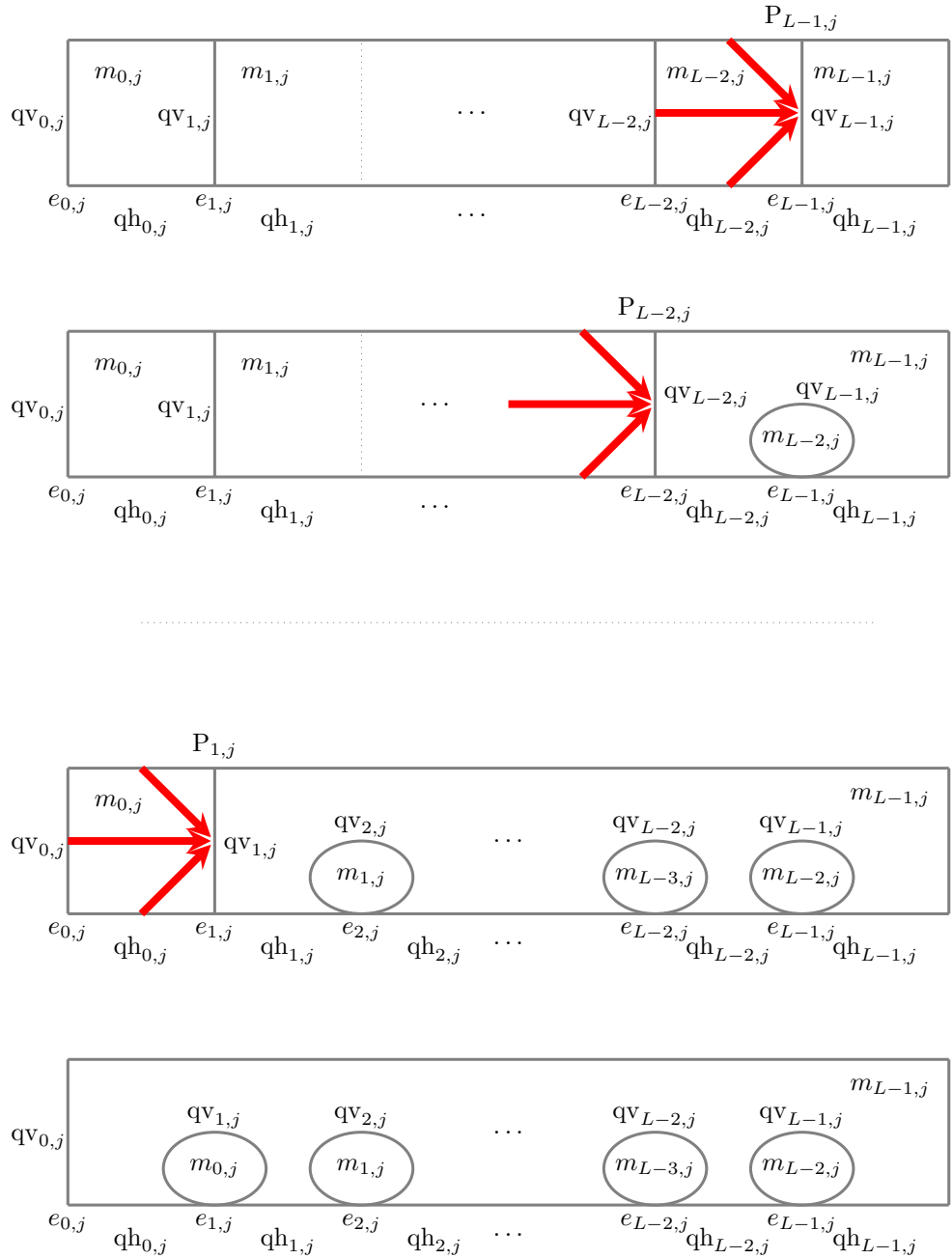}
\caption{\label{fig:fullstructurerows} {\small \textbf{General toric
      code simplification, step 1.} Creating bubbles along each row by
    P-moves from right to left.}}
\end{figure}

Figure \ref{fig:fullstructurerows}.

Operation $\mathcal{U}_1 = \bigotimes_{j=0}^{T-1} U^{\text{row (1)}}_j$,
with $U^{\text{row (1)}}_j = \mathrm{P}_{1, \, j} \mathrm{P}_{2, \, j}
\ldots \mathrm{P}_{L-1, \, j}$.  P-move $\mathrm{P}_{i, \, j}$ is
defined as the application of CNOTs on target qubit $\mathrm{qv}_{i,
  \, j}$ with controls the rest of the qubits of plaquette
$\mathrm{p}_{i-1, \, j}$.  Explicitly,
\begin{equation}\label{bubbles:defpmovep}
  \mathrm{P}_{i, j}
=
  \mathcal{C} (
    \mathrm{qv}_{i, j} \leftarrow \mathrm{qh}_{i-1, \, j+1}
  )
  \mathcal{C} (
    \mathrm{qv}_{i, j} \leftarrow \mathrm{qv}_{i-1, \, j}
  )
  \mathcal{C} (
    \mathrm{qv}_{i, j} \leftarrow \mathrm{qh}_{i-1, \, j}
  )
  \; .
\end{equation}

\paragraph{\label{para:rowsbis}%
  Step 2: Create spikes along rows}

\begin{figure}[htbp]
\centering
\includegraphics{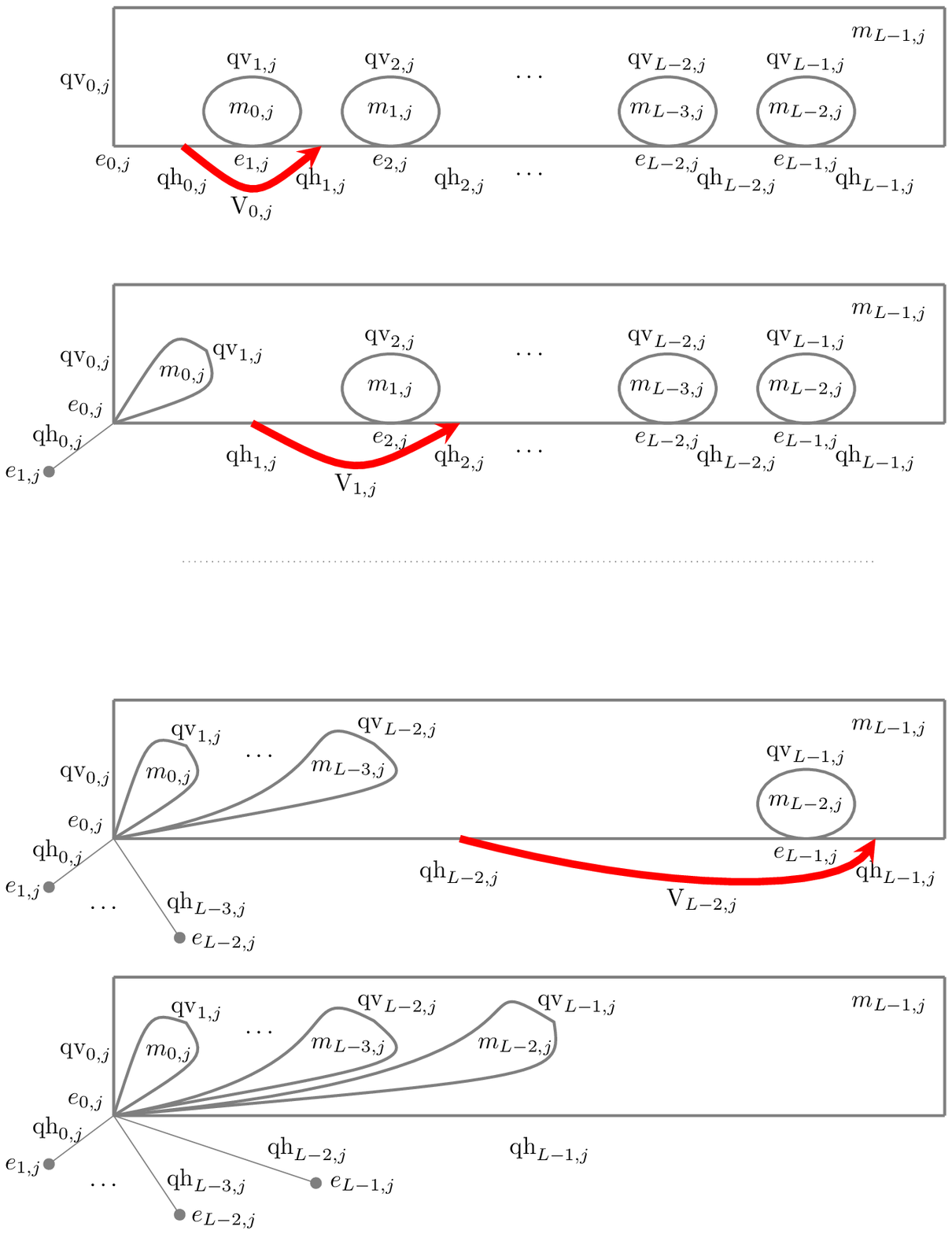}
\caption{\label{fig:fullstructurerowsbis} {\small \textbf{General
      toric code simplification, step 2.} Creating spikes along each
    row by V-moves from left to right.}}
\end{figure}

Figure \ref{fig:fullstructurerowsbis}.

Operation $\mathcal{U}_2 = \bigotimes_{j=0}^{T-1} U^{\text{row
    (2)}}_j$, with $U^{\text{row (2)}}_j = \mathrm{V}_{L-2, \, j}
\mathrm{V}_{L-3, \, j} \ldots \mathrm{V}_{0, \, j}$.  Each V-move here
consists of a single CNOT, $\mathrm{V}_{i, \, j} = \mathcal{C} (
\mathrm{qh}_{i+1, j} \leftarrow \mathrm{qh}_{i, \, j} )$, since the
qubit in the affected bubble ($\mathrm{qv}_{ i+1, \, j }$) is acted
upon twice, and hence unchanged.

At this stage, the toric lattice has been simplified to a single
column, to the vertices of which are attached a collection of bubbles
and spikes.  The next moves reproduce the strategy followed before for
each row.

\paragraph{\label{para:column}%
  Step 3: Create bubbles along the column}

\begin{figure}[htbp]
\centering
\includegraphics{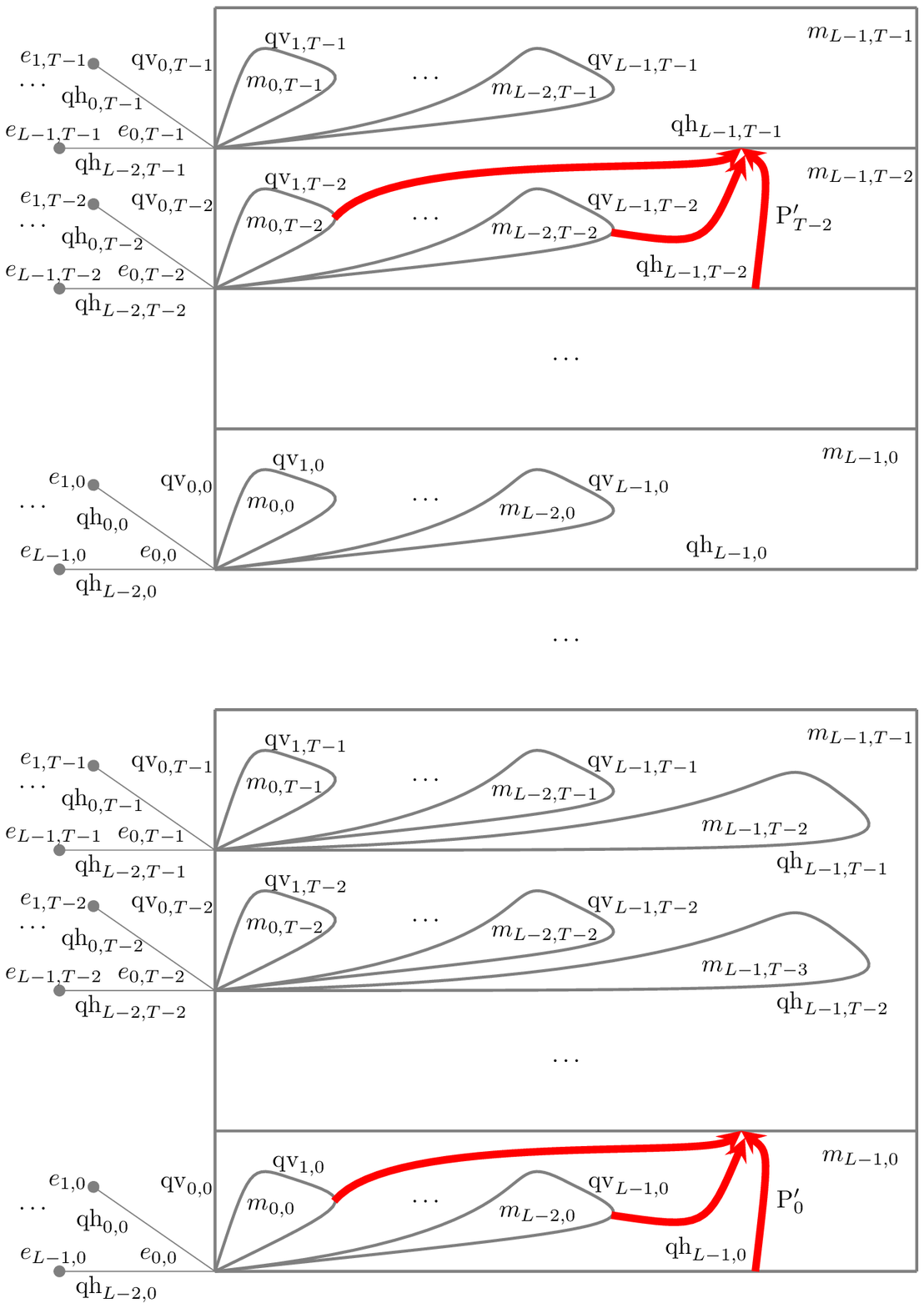}
\caption{\label{fig:fullstructurecolumn} {\small \textbf{General toric
      code simplification, step 3.} Creating bubbles along the
    remaining column by P-moves from top to bottom.}}
\end{figure}

Figure \ref{fig:fullstructurecolumn}.

Operation $\mathcal{U}_3 = \mathrm{P}'_0 \ldots \mathrm{P}'_{T-2}$,
where each P-move consists now of $L$ CNOTs,
\begin{align}\label{bubbles:defpmovepprime}
\nonumber
  \mathrm{P}'_j
&=
  \mathcal{C} (
    \mathrm{qh}_{L-1, \, j+1} \leftarrow \mathrm{qv}_{1, \, j}
  )
  \ldots
\\
&
\quad
\times
  \mathcal{C} (
    \mathrm{qh}_{L-1, \, j+1} \leftarrow \mathrm{qv}_{L-1, \, j}
  )
  \mathcal{C} (
    \mathrm{qh}_{L-1, \, j+1} \leftarrow \mathrm{qh}_{L-1, \, j}
  )
  \; .
\end{align}

\paragraph{\label{para:columnbis}%
  Step 4: Create spikes along the column}

\begin{figure}[htbp]
\centering
\includegraphics{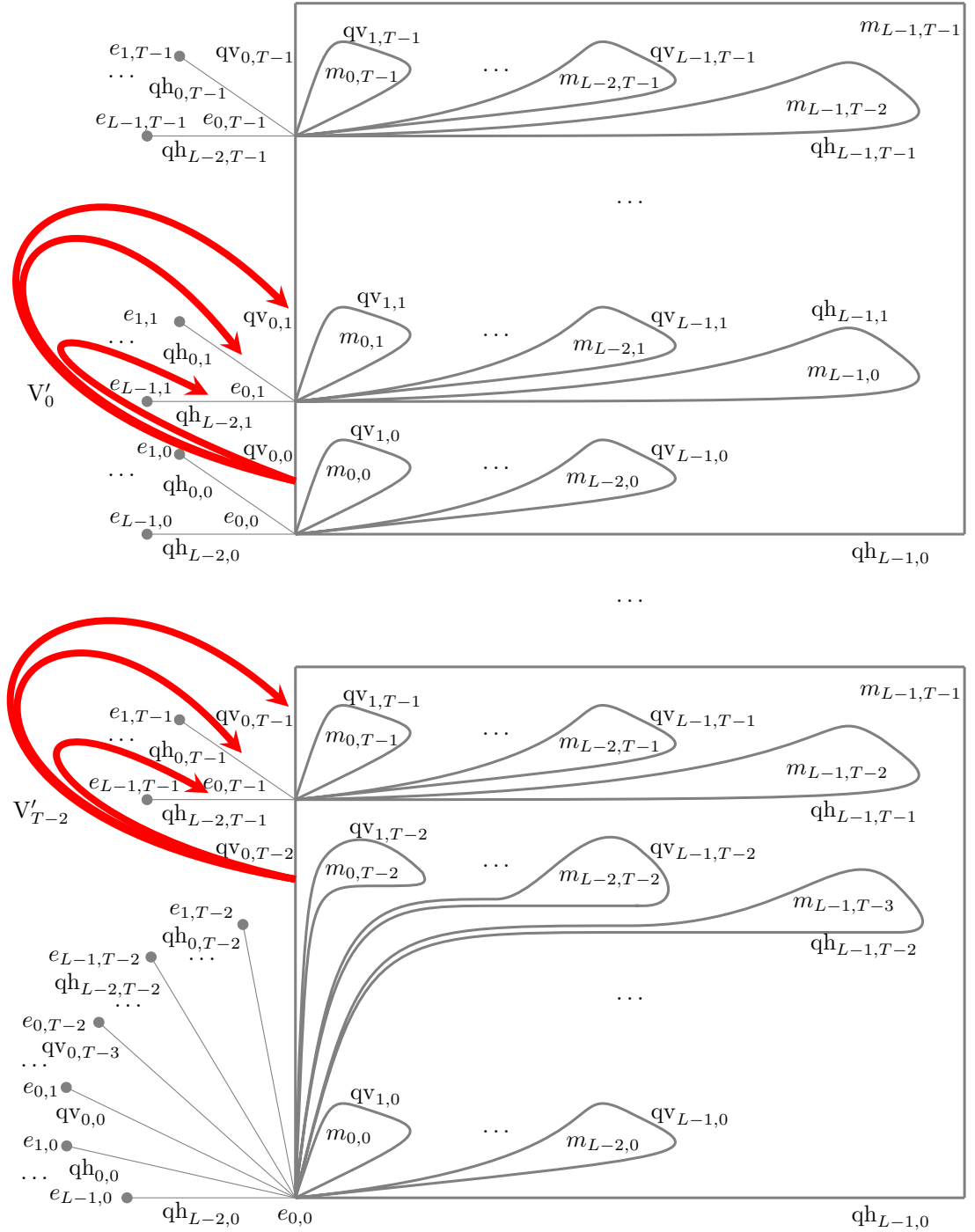}
\caption{\label{fig:fullstructurecolumnbis} {\small \textbf{General
      toric code simplification, step 4.} Creating spikes along the
    column by V-moves from bottom to top.}}
\end{figure}

Figure \ref{fig:fullstructurecolumnbis}.

Operation $\mathcal{U}_4 = \mathrm{V}'_{T-2} \ldots \mathrm{V}'_0$,
where each V-move consists now of $L$ CNOTs,
\begin{equation}\label{bubbles:defvmove}
  \mathrm{V}'_j
=
  \mathcal{C} (
    \mathrm{qh}_{L-2, \, j+1} \leftarrow \mathrm{qv}_{0, \, j}
  )
  \ldots
  \mathcal{C} (
    \mathrm{qh}_{0, \, j+1} \leftarrow \mathrm{qv}_{0, \, j}
  )
  \mathcal{C} (
    \mathrm{qv}_{0, \, j+1} \leftarrow \mathrm{qv}_{0, \, j}
  )
  \; .
\end{equation}

\begin{figure}[htbp]
\centering
\includegraphics{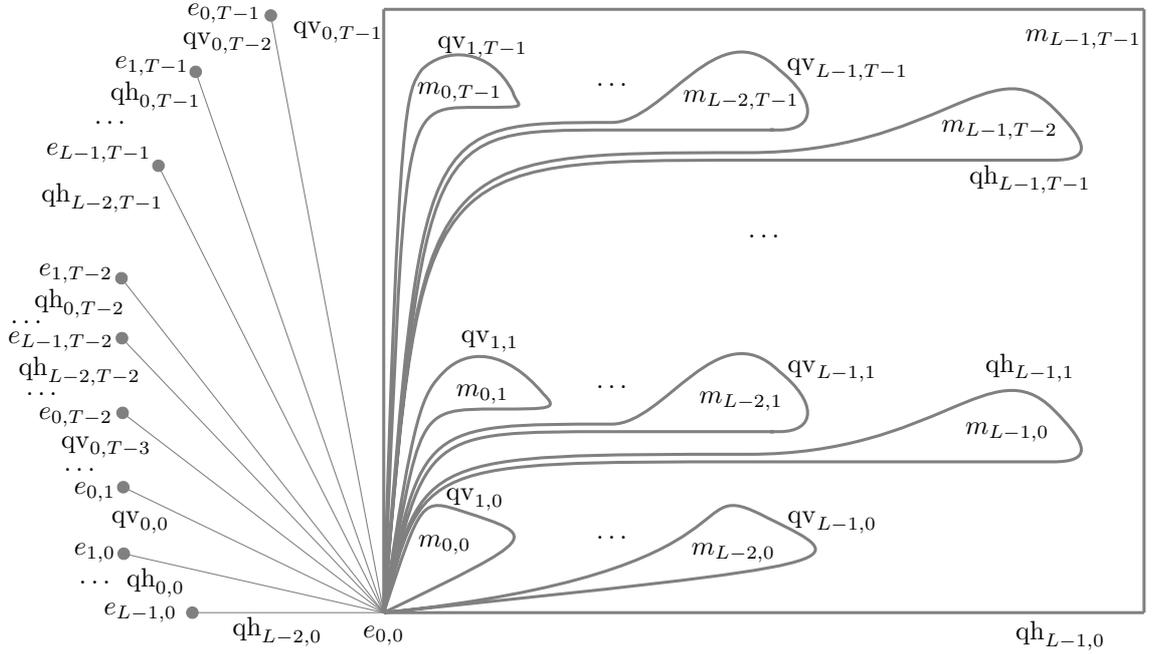}
\caption{\label{fig:fullstructurefinal} {\small \textbf{General toric
      code simplification, step 5.} Final lattice.}}
\end{figure}

After this set of operations, the topological information is carried
by qubits $\mathrm{qh}_{L-1, \, 0}$ (eigenvalue $z_H$) and
$\mathrm{qv}_{0, \, T-1}$ (eigenvalue $z_V$).  All magnetic fluxes in
the original plaquettes except for $m_{L-1, \, T-1}$ are now located
in one-qubit bubbles, according to the rule
\begin{align}\label{bubbles:ruleflux}
\nonumber
&
  \text{%
    flux $m_{i, \, j}$ is carried by
    qubit $\mathrm{qv}_{i+1, \, j}$,
    $0 \leq i \leq L-2$, $\forall j$,}
\\
&
  \text{%
    flux $m_{L-1, \, j}$ is carried by
    qubit $\mathrm{qh}_{L-1, \, j+1}$,
    $\forall j$.}
\end{align}
Flux $m_{L-1, \, T-1}$ ensures overall flux neutrality and is carried
by the large plaquette on whose perimeter lie all qubits (with the
caveat that qubits carrying topological information appear twice, so
do not contribute to the associated stabiliser).  Similarly, all
electric charges except for $e_{0, \, 0}$ appear now in vertices at
the ends of one-qubit spikes, according to the rule
\begin{align}\label{bubbles:rulecharge}
\nonumber
&
  \text{%
    charge $e_{0, \, j}$ is carried by
    qubit $\mathrm{qv}_{0, \, j-1}$,
    $1 \leq j \leq T-1$,} \\
&
  \text{%
    charge $e_{i, \, j}$ is carried by
    qubit $\mathrm{qh}_{i-1, \, j}$,
    $1 \leq i \leq L-1$, $\forall j$,}
\end{align}
and charge $e_{0, \, 0}$ appears at a vertex shared by all qubits
(twice by both qubits carrying topological information).  The final
lattice is drawn in figure \ref{fig:fullstructurefinal}.

Hence, the operation $\mathcal{U}_4 \mathcal{U}_3 \mathcal{U}_2
\mathcal{U}_1$ diagonalises the Hamiltonian in the sense that (a)
logical information does not appear in the Hamiltonian but is encoded
in two physical qubits, (b) all independent eigenvalues of stabilisers
in the Hamiltonian appear now as eigenvalues of one-body operators.
Global topological charge neutrality is ensured by two many-body
terms.  As explained before, a transformation into two copies of an
Ising chain can subsequently be performed.

As regards the efficiency of the procedure, the number of CNOTs to be
effected is $\mathcal{O} (LT)$. If operations are parallelised as much
as possible, the operation can be performed in $\mathcal{O} ( L + T )$
steps.

\clearpage

\section{\label{sec:qdoubles}%
  Quantum double models}

In this section we apply the ER-inspired scheme described in the
Abelian case of the toric code to general quantum double models
$\mathrm{D} (G)$ based on non-Abelian groups \cite{Kitaev:1997}.
While this construction holds exactly, with a more natural structure,
in the class of quantum double models $\mathrm{D} (H)$ based on Hopf
algebras \cite{BMCA}, which contains and generalises the $\mathrm{D}
(G)$ models, we settle for a discussion of the group case since the
language is bound to be more familiar.

\subsection{\label{subsec:qdoubles:hamilt}%
  Hamiltonian, charges}

The quantum double model based on a finite group $G$, or $\mathrm{D}
(G)$ model, is defined on an 2D lattice $\Lambda$ where quantum
degrees of freedom of dimension $\lvert G \rvert$ sit on the edges.
Orthonormal bases $\{ \lvert g \rangle \}$ are chosen for each
oriented edge, labelled by group elements; the bases for the two
orientations of any given edge are related by the inversion $g \mapsto
g^{-1}$.  The Hilbert space of an oriented edge may be identified with
the group algebra $\mathbb{C} G$ of complex combinations of group
elements.

The Hamiltonian of the $\mathrm{D} (G)$ model is
\begin{equation}\label{eq:hamiltoniandgmodel}
  H_\Lambda
=
 - \sum_P B_P - \sum_V A_V
\; ,
\end{equation}
where the first sum runs over plaquettes of $\Lambda$, the second over
vertices of $\Lambda$, and $A_V$, $B_P$ are mutually commuting
projectors with support on the edges adjacent to the vertex and along
the boundary of the plaquette, respectively.

\begin{figure}[htbp]
\centering
\includegraphics{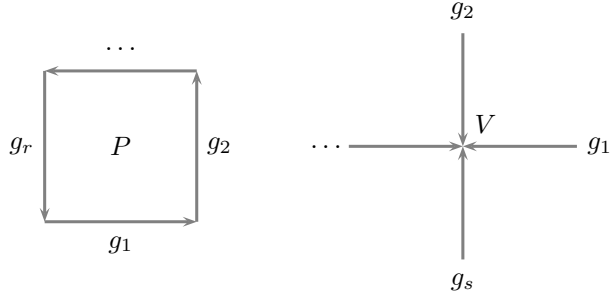}
\caption{\label{fig:conventionshdg} {\small \textbf{Quantum double
      Hamiltonian.} Conventions for the definition of plaquette and
    vertex projectors in equations (\ref{eq:defplaquetteprojectors})
    and (\ref{eq:defvertexprojectors}).}}
\end{figure}

Plaquette projectors select configurations where the product of group
elements along the plaquette boundaries is the identity element of
$G$:
\begin{equation}\label{eq:defplaquetteprojectors}
  B_P \,
  \lvert g_1, \, g_2, \, \ldots, \, g_r \rangle
=
  \delta_e ( g_r \ldots g_2 g_1 ) \,
  \lvert g_1, \, g_2, \, \ldots, \, g_r \rangle
\; ,
\end{equation}
where $g_1, \, \ldots, \, g_r$ are group elements defining a
computational basis state of the edges along the plaquette boundary,
oriented and ordered along an anticlockwise circuit with arbitrary
origin, as illustrated in figure \ref{fig:conventionshdg}.  Vertex
projectors have the form
\begin{equation}\label{eq:defvertexprojectors}
  A_V \,
  \lvert g_1, \, g_2, \, \ldots, \, g_s \rangle
=
  \frac{ 1 }{ \lvert G \rvert } \, 
  \sum_{ k \in G }
  \lvert k g_1, \, k g_2, \, \ldots, \, k g_s \rangle
\; ,
\end{equation}
where $g_1, \, \ldots, \, g_s$ are group elements defining a
computational basis ket in the bonds adjacent to the vertex, oriented
towards it, as in figure \ref{fig:conventionshdg}.

Simultaneous $+1$ eigenstates of all $A_V$ and $B_P$ are ground
states, and excited states break the ground state conditions $A_V
\lvert \psi \rangle = \lvert \psi \rangle$, $B_P \lvert \psi \rangle =
\lvert \psi \rangle$ at some vertices or plaquettes.  We can speak of
particle-like excitations living at broken plaquettes and vertices,
and it turns out that these particles are characterised by topological
charges from an anyon model \cite{Kitaev:2005}.  Topological charges
are properties of regions that cannot be changed by physical
operations within the region, and can be determined from measurements
on its boundary (think of electric charge, measured by Gauss's law);
more generally, a topological charge label can be assigned to any
closed loop independently of whether it is the boundary of a region or
not, which is crucial for the study of topologically nontrivial
systems, as we will see in the next section.

Topological charges in the quantum double models can be understood
algebraically: they are labelled by irreducible representations of
Drinfel'd's \textsl{quantum double} $\mathrm{D} (G)$, a
quasitriangular Hopf algebra which can be computed for each group $G$
\cite{Drinfeld}.  More details will be given later; for the moment let
us remark that the distribution of topological charge on the lattice
may be characterised locally by assigning charge labels to lattice
\textsl{sites}, where, following Kitaev, a site is a pair of a
plaquette and one of its neighbouring vertices.  A charge has thus in
general a magnetic part, associated with the plaquette, and an
electric part, associated with the vertex; a charge with both
nontrivial magnetic and electric parts is called a dyon.  The part of
Hamiltonian (\ref{eq:hamiltoniandgmodel}) acting on a site assigns
energy $-2$ to the trivial charge (the unique charge with trivial
electric and magnetic parts), $-1$ to purely magnetic or purely
electric charges, and $0$ to dyons.

\subsection{\label{subsec:qdoubles:smallesttorus}%
  The smallest torus}

To generalise the treatment of the toric code given in section
\ref{sec:toric}, we first need to understand the topological degrees
of freedom of the $\mathrm{D} (G)$ model on the torus, which are not
covered by the previous \textsl{local} characterisation of topological
charge distributions.

\begin{figure}[htbp]
\centering
\includegraphics{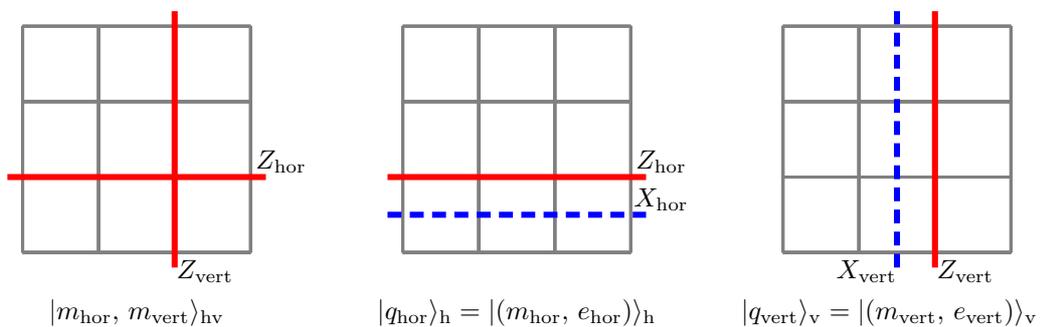}
\caption{\label{fig:smalltorusdg} {\small \textbf{Logical operators in
      the toric code}. Sets of commuting observables yielding complete
    topological information in the ground level. \textsl{Left}: Two
    logical Pauli $Z$s measuring `half the topological charge' along
    each noncontractible loop on $\mathrm{T}^2$. \textsl{Centre}: A
    pair of logical Pauli $Z$ and $X$ measuring the whole topological
    charge along the horizontal loop.  \textsl{Right}: The same for
    the vertical loop.  In each case, the associated bases for the
    ground level are displayed.}}
\end{figure}

In the toric code, the ground level has fourfold degeneracy; usually
this is interpreted as the set of states of two logical qubits
associated with the two noncontractible loops on $\mathrm{T}^2$ as in
the left part of figure \ref{fig:smalltorusdg}.  The logical Pauli
$Z_{\mathrm{hor}}$ and $Z_{\mathrm{vert}}$ measure \textsl{magnetic}
quantum numbers $m_{\mathrm{hor}}$, $m_{\mathrm{vert}}$ along the
horizontal and vertical loops; this amounts to determining `half the
topological charge' for each loop.  This defines a basis $\lvert
m_{\mathrm{hor}}, \, m_{\mathrm{vert}} \rangle_{\mathrm{hv}}$ of
eigenstates in the ground level.

Alternatively, one may choose as commuting set of observables the
logical Pauli $Z_{\mathrm{hor}}$ along a horizontal loop of the direct
lattice, and the logical Pauli $X_{\mathrm{hor}}$ along a horizontal
loop of the dual lattice (centre of figure \ref{fig:smalltorusdg}).
In this case one has a basis $\lvert q_{\mathrm{hor}}
\rangle_{\mathrm{h}}$ labelled by the \textsl{full} topological charge
label $q_{\mathrm{hor}} = ( m_{\mathrm{hor}}, \, e_{\mathrm{hor}} )$
associated with the horizontal loop, including the eigenvalue
$m_{\mathrm{hor}}$ of the magnetic $Z_{\mathrm{hor}}$ operator and the
eigenvalue $e_{\mathrm{hor}}$ of the electric $X_{\mathrm{hor}}$
operator.
 
If we choose the corresponding logical Pauli operators along the
vertical loop (figure \ref{fig:smalltorusdg}, right), then we have
another basis $\lvert q_{\mathrm{vert}} \rangle_{\mathrm{v}}$ labelled
by a full topological charge.  From the abstract properties of an
anyon model \cite{Kitaev:2005}, we know that the change of basis
between $\lvert q_{\mathrm{hor}} \rangle_{\mathrm{h}}$ and $\lvert
q_{\mathrm{vert}} \rangle_{\mathrm{v}}$ is the unitary operator known
as the topological $S$-matrix:
\begin{equation}\label{eq:toposmatrix}
  \lvert q_{\mathrm{vert}} \rangle_{\mathrm{v}}
=
  \sum_{ q_{\mathrm{hor}} }
  \lvert q_{\mathrm{hor}} \rangle_{\mathrm{h}} \,
  S_{ q_{\mathrm{hor}}, \, q_{\mathrm{vert}} }
\; .
\end{equation}
This relation encodes a sort of uncertainty principle for topological
labels on $\mathrm{T}^2$: if we determine the full label along the
horizontal loop, the vertical label is `maximally spread', and vice
versa.  The states of the $\mathrm{hv}$-basis work as sorts of
coherent states where a compromise in the measurements of the
horizontal and vertical labels is made so as to have mutually
commuting observables.

When considering non-Abelian models, we should not expect the
$\mathrm{hv}$-basis to survive.  Indeed, in an Abelian $\mathrm{D}
(G)$ model the number of topological sectors is $\lvert G \rvert^2$, a
perfect square, and one can successfully construct the $\lvert
m_{\mathrm{hor}}, \, m_{\mathrm{vert}} \rangle_{\mathrm{hv}}$ states
due to the separation of electric and magnetic degrees of freedom.  In
a non-Abelian $\mathrm{D} (G)$ model one has in general a number of
charges different from a perfect square, for instance eight charges in
the $\mathrm{D} ( \mathrm{S}_3 )$ model.  

However, the $\mathrm{h}$-basis and the $\mathrm{v}$-basis can be
generalised for non-Abelian models.  More than that, their existence
stems from the modular nature of the underlying anyon model and is
therefore a fundamental property.

As in the Abelian case, a quantum double model on the minimal toric
lattice consisting of two edges contains the full basis of the ground
level; in contrast with the Abelian case, the Hilbert space of these
two qudits is not identical with the ground level, that is, there is
room for excitations even in such a small system.  Exactly for
\textsl{which} excitations is ultimately a representation-theoretical
problem.

\begin{figure}[htbp]
\centering
\includegraphics{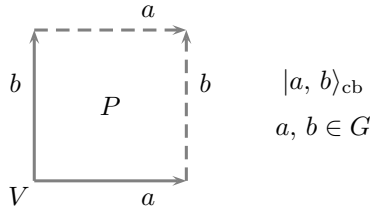}
\caption{\label{fig:smalltorusdgbasis} {\small \textbf{Computational
      basis for the minimal torus in the $\mathrm{D} (G)$ model.}}}
\end{figure}
\begin{figure}[htbp]
\centering
\includegraphics{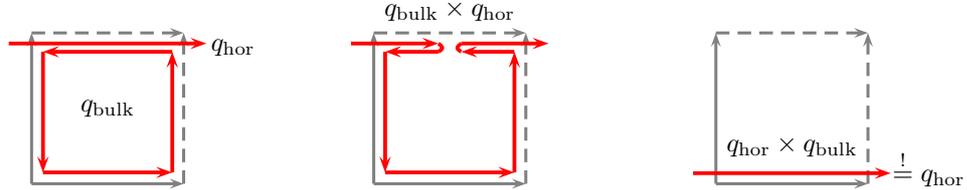}
\caption{\label{fig:consistencyconditioncharges} {\small
    \textbf{Consistency condition for loop and bulk charges.}  The
    loop in the right hand figure is, by periodicity, equivalent to
    the first horizontal loop, so the charge measured there has to be
    the same.  Since the final loop is obtained by concatenating the
    original loop and the circuit measuring the bulk charge, the
    condition $q_{ \mathrm{hor} } \times q_{ \mathrm{bulk} }
    \rightarrow q_{ \mathrm{hor} } + \ldots$ follows.}}
\end{figure}

To wit, let $\lvert a, \, b \rangle_{\mathrm{cb}}$ be the elements of
the computational basis for the minimal torus, with $a, \, b \in G$
(figure \ref{fig:smalltorusdgbasis}).  The action of the unique
plaquette and vertex projectors reads
\begin{align}\label{eq:smalltorusavbp}
\nonumber
  B_P \,
  \lvert a, \, b \rangle_{\mathrm{cb}}
&=
  \delta ( ab, \, ba ) \,
  \lvert a, \, b \rangle_{\mathrm{cb}} \; ,
\\
  A_V \,
  \lvert a, \, b \rangle_{\mathrm{cb}}
&=
  \frac{ 1 }{ \lvert G \rvert } \,
  \sum_{ g \in G }
  \lvert
    g a g^{-1}, \, g b g^{-1}
  \rangle_{\mathrm{cb}} \; ,
\end{align}
Clearly, if $G$ is Abelian these are always the identity operator, so
the Hilbert space of the two qudits is identical with the ground
level.  But for non-Abelian groups, the conditions $B_P \, \lvert \psi
\rangle = \lvert \psi \rangle$, $A_V \, \lvert \psi \rangle = \lvert
\psi \rangle$ are nontrivial.  We can specify a state by giving two
sets of charge information:
\begin{itemize}
\item %
  A horizontal charge label $q_{\mathrm{hor}}$ corresponding to one
  horizontal nontrivial loop, as in the previous discussion.
\item %
  A bulk charge label $q_{\mathrm{bulk}}$ corresponding to the charge
  sitting at the \textsl{site} defined by the unique plaquette and the
  unique vertex of the minimal square, together with internal degrees
  of freedom of the bulk charge as dictated by the microscopic model.
\end{itemize}
Not all pairs $( q_{\mathrm{hor}}, \, q_{\mathrm{bulk}} )$ are
compatible: in particular, some charges never appear in the bulk,
while some bulk charges are incompatible with some, but not all, of
the loop charge labels.  The condition for such a pair to be
compatible is that the fusion of the two charges contains the
horizontal charge:
\begin{equation}\label{eq:consistencybulkhorcharges}
  q_{\mathrm{hor}} \times q_{\mathrm{bulk}}
\rightarrow
  q_{\mathrm{hor}} + \ldots
\end{equation}
The reason for this is consistency of the charge measurement along the
loop: if the loop is deformed by sweeping the whole bulk of the torus,
the same charge label should be measured in the resulting loop,
because of periodicity.  Since the new loop is equivalent to the
concatenation of the original loop and a loop enclosing the bulk
charge, condition (\ref{eq:consistencybulkhorcharges}) follows (see
figure \ref{fig:consistencyconditioncharges}).

The discussion of these phenomena for $\mathrm{D} (G)$ models requires
some algebraic machinery and we leave it for appendix
\ref{sec:appendix:dgmodels}, where we attempt a pedagogical account;
the case of the smallest non-Abelian group $\mathrm{S}_3$ is
considered in appendix \ref{sec:appendix:dsthree}.  Here we just list
the results:
\begin{itemize}
\item %
  Charges in a $\mathrm{D} (G)$ model are labelled by irreducible
  representations of Drinfel'd's quantum double algebra $\mathrm{D}
  (G)$.  These are given by pairs $( C, \, \alpha )$, where $C$ are
  conjugacy classes of $G$, and $\alpha$ are irreducible
  representations of the centraliser group of $C$; more details appear
  in appendix \ref{sec:appendix:dgmodels}.
\item %
  The Hilbert space in the smallest torus consisting of two edges has
  an orthonormal basis:
\begin{equation}\label{eq:basissmalltorus}
  \big\{
    \lvert
      (C, \alpha; k, \, v_i )_{ \mathrm{bulk} };
      (C', \alpha')_{ \mathrm{hor} }
    \rangle ; \ {}
    ( C, \alpha ) \otimes ( C', \, \alpha' )
   \rightarrow
    ( C', \, \alpha' ) \oplus \ldots
  \big\}
\; ,
\end{equation}
  where $k, \, v_i$ are degrees of freedom determining specific
  vectors in an irreducible representation space for $( C, \, \alpha
  )$.  The projectors onto definite bulk and horizontal charge labels
  are given in equation (\ref{eq:ribbonprojectors}) in appendix
  \ref{sec:appendix:dgmodels}.
\item %
  From the point of view of the Hamiltonian of the $\mathrm{D} (G)$
  model, configurations with vacuum bulk charge $( C_e, \, 1_+ )$
  (where the conjugacy class is $C_e = \{ e \}$, and $1_+$ is the
  trivial representation of $G$) are ground states, with energy $-2$.
  The energy of states with pure magnetic charges $( C, \, 1_+ )$
  (where $C \neq C_e$, and $1_+$ is the trivial representation of the
  centraliser of $C$) and that of pure electric charges $( C_e, \alpha
  )$ (where $\alpha \neq 1_+$) in the bulk is $-1$, due to the
  breakdown of the vertex and plaquette condition, respectively.
  Finally, the energy of states with a bulk charge $( C, \, \alpha )$
  that breaks both the plaquette and vertex conditions ($C \neq C_e$,
  $\alpha \neq 1_+$) is $0$.
\end{itemize}

\subsection{\label{subsec:qdoubles:elementary}%
  Elementary moves}

The generalisation of the Abelian P-moves and V-moves to non-Abelian
quantum double models was given in the Appendix of the preprint
version of \cite{AguadoVidal}, in the case of groups; essentially, the
CNOT operations of the Abelian case turn into controlled
multiplication operators of the type $\lvert g, \, h \rangle \mapsto
\lvert g, \, g h \rangle$ and their variants obtained by changes of
orientation of the edges for control and target qudits, and these
operators are composed to build the elementary moves.  For models
based on Hopf algebras, the elementary moves, also constructed from
controlled multiplications, were defined in \cite{BMCA} and give rise
to a MERA for the ground states and the disentangling procedure of
this paper in an analogous way; but as already mentioned, we
concentrate on the case of groups here.  We will not give general
expressions, but concentrate on simple examples from which the general
procedure is straightforwardly inferred.

\begin{figure}[htbp]
\centering
\includegraphics{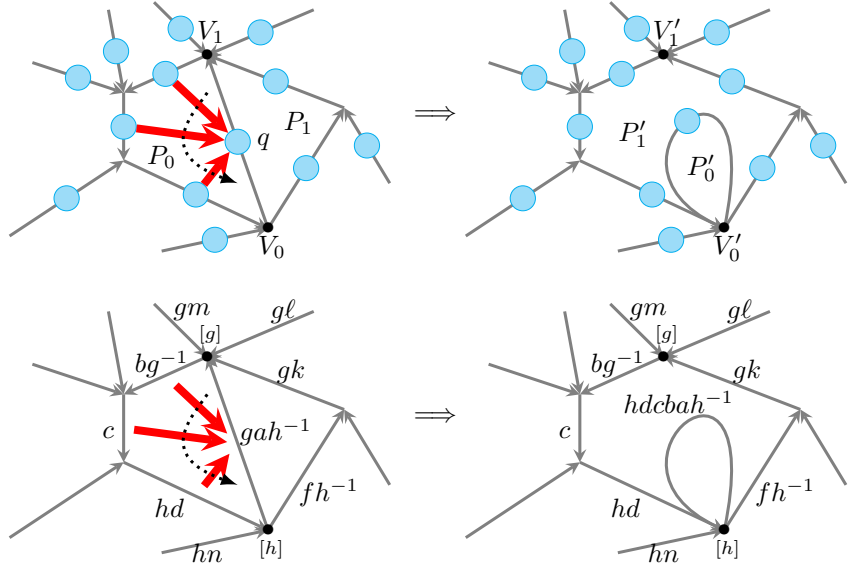}
\caption{\label{fig:dgpmove} {\small \textbf{P-moves for a general
      $\mathrm{D} (G)$ model.}  Solid arrows stand for controlled left
    group multiplication operators in the form $\lvert g, \, h \rangle
    \mapsto \lvert g, \, g h \rangle$; these are to be applied in the
    order defined by the dotted arrow, assuming the orientation of the
    edges involved as in the figure.  All labels are group elements;
    therefore we show kets in the computational basis.  We give the
    action of vertex representations of the group (gauge
    transformations) by showing how elements $h$, $g$ are represented
    in the vertices $V_0$, $V_1$ at the boundary of the target edge.
    After the P-move, the result is clearly a quantum double model in
    the new lattice that we draw, where the target edge talks twice to
    the lowermost vertex $V'_0$ and no longer has contact with the
    uppermost vertex $V'_1$.  This we call a P-move associated with
    site $(P_0, \, V_0)$.}}
\end{figure}

A plaquette move or P-move has to be specified more carefully than in
the Abelian case, since controlled multiplication operators acting on
a target qudit no longer commute with each other.  In general, we need
specify both a plaquette and one of its edges as a target, or
equivalently, one of the vertices along the boundary of the plaquette.
A convention necessarily links the specific multiplication operators
and the lattice transformation, including the orientation of the
edges: ours is defined in figure \ref{fig:dgpmove}, which we choose to
associate with the site $(P_0, \, V_0)$.

\begin{figure}[htbp]
\centering
\includegraphics{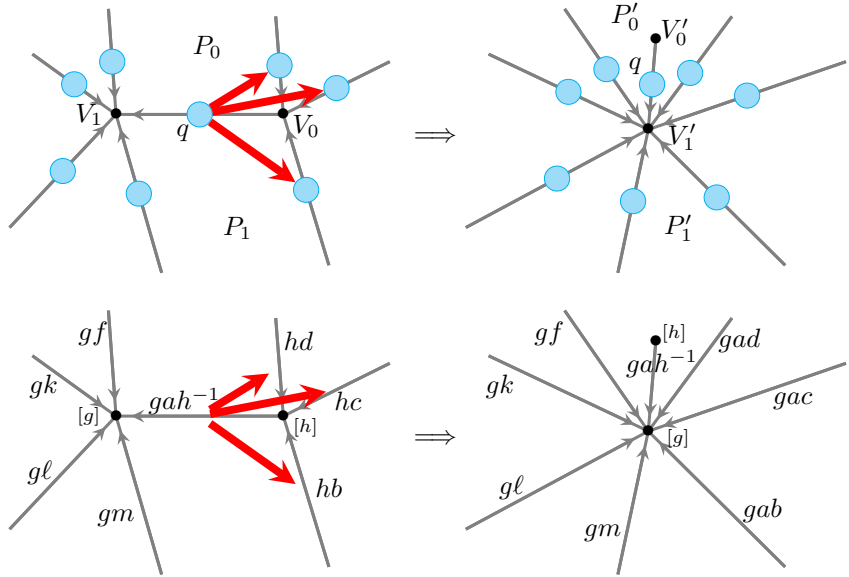}
\caption{\label{fig:dgvmove} {\small \textbf{V-moves for a general
      $\mathrm{D} (G)$ model.}  Solid arrows, as before, stand for
    controlled multiplications $\lvert g, \, h \rangle \mapsto \lvert
    g, \, g h \rangle$; here they need not be applied in any
    particular order.  The V-move in the computational basis, as well
    as gauge transformations at the affected vertices, are shown in
    the bottom row.  After the move, the control edge is the only one
    in contact with vertex $V'_0$, and therefore the only edge feeling
    the gauge transformation by $h$.  The plaquette in which the
    `spike' is drawn is arbitrary from the point of view of the
    Hamiltonian; but to be definite we associate this V-move with site
    $(V_0, \, P_0)$.}}
\end{figure}

The vertex moves or V-moves are defined analogously, with the
difference that one uses controlled multiplications with a single
control and multiple targets, that commute with each other, so one
does not need to apply these operations in a certain order.  In the
models based on Hopf algebras, this commutativity property is lost and
one has to take care of the order of operations as well \cite{BMCA}
(the property that makes the group case simpler is the cocommutativity
of the Hopf structure in group algebras).  Our conventions are defined
in figure \ref{fig:dgvmove}, illustrating a V-move which we associate
with the site $(V_0, \, P_0)$.

\subsection{\label{subsec:qdoubles:disenthamilt}%
  Disentangling quantum double models}

For the purpose of analysing the quantum double Hamiltonian
(\ref{eq:hamiltoniandgmodel}), the only relevant questions are those
asked by the vertex and plaquette projectors $A_v$, $B_p$, whose
eigenvalues we may consider a distribution of ``binary charges''
$e_v$, $m_p$ taking values in $\{ 0, \, 1 \}$.  As in the Abelian
case, the original Hamiltonian gets transformed into the quantum
double Hamiltonian in the deformed lattice, and the ``binary charges''
get reassigned as depicted in figures
\ref{bubbles:fig:recoupleplaqflux} and
\ref{bubbles:fig:recouplevertflux}.  The geometrical moves conserve
the number of edges (degrees of freedom), vertices and plaquettes, and
the operations on the state are unitaries preserving the energy.

The disentangling procedure for general quantum double Hamiltonians
(\ref{eq:hamiltoniandgmodel}) has thus the same graph-theoretical
structure as the Abelian cases.  The key is that the elementary moves
simply redistribute the vertex and plaquette projectors along with the
local deformations of the lattice, according to the general rule
\begin{equation}\label{eq:disentanglegeneralruleh}
  U_{ P (V) } \,
  H_\Lambda \, 
  U_{ P (V) }^\dagger
=
  H_{ \Lambda'_{ P (V) } } \, 
\; ,
\end{equation}
where $\Lambda$ is the original lattice, and $\Lambda'_{ P (V) }$ are
the deformed lattices obtained after the P- and V-moves.

If we work on a topologically nontrivial surface, there appears a
difference with the Abelian models.  Consider again the case of the
torus: eventually the situation simplifies into a lattice where just
two ``topological'' edges go around the two nontrivial cycles, and
only one plaquette and one vertex talk (twice each) to them.  In the
Abelian case the topological edges do not contribute to the dynamics,
essentially because the bulk charge in the quantum double model on the
small torus consisting of just these two edges, as discussed
previously, is always trivial.  In a non-Abelian case, the small torus
may contain bulk charges, as constrained by condition
(\ref{eq:consistencybulkhorcharges}), and these contribute to the
energy of the disentangled quantum double model.

More precisely, let us consider the two simple cases analogous to the
disentangling of the toric code into mostly sums of one-body operators
(figure \ref{fig:twobytwocleaner}), and the mapping to the double
Ising spin chain (figure \ref{fig:twobytwotoising}).

\begin{figure}[htbp]
\centering
\includegraphics{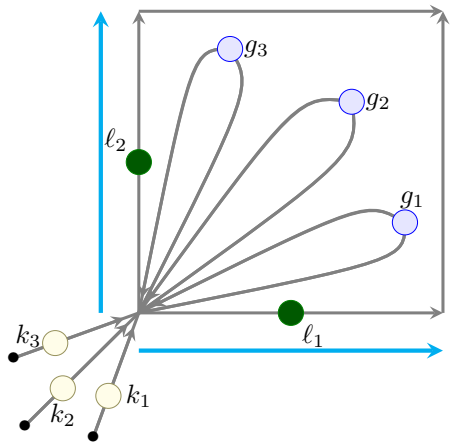}
\caption{\label{fig:dgonebodyterms} {\small \textbf{Disentangled
      quantum double.}  The lattice for the expression
    (\ref{eq:hamiltoniandgonebody}) of the Hamiltonian mostly as a sum
    of one-body terms.  In a ket in the computational basis,
    `magnetic' qudits are in states $\lvert g_i \rangle$, `electric
    qudits' are in states $\lvert k_j \rangle$, and topological qudits
    are in states $\lvert \ell_1 \rangle$, $\lvert\ell_2 \rangle$.}}
\end{figure}
In the first case, illustrated in figure \ref{fig:dgonebodyterms}, the
qudits separate into three classes: $r$ `magnetic qudits', whose edges
enclose a whole plaquette; $s$ `electric' qudits, whose edges end at a
vertex with no other edges attached to it; and two topological qudits.
We write a ket in the computational basis as $\lvert g_1, \, \ldots,
\, g_r; \, k_1, \, \ldots, k_s; \, \ell_1, \, \ell_2 \rangle$
according to the orientation conventions of figure
\ref{fig:dgonebodyterms} (where $r = s = 3$), where the $\{ g_i \}$,
$\{ k_j \}$ and $\ell_1, \, \ell_2$ correspond to the magnetic,
electric, and topological qudits, respectively.  The Hamiltonian reads
\begin{equation}\label{eq:hamiltoniandgonebody}
  H
=
 - \, \sum_{ i = 1 }^r B_i
 - \, B_0
 - \, \sum_{ j = 1 }^s A_j
 - \, A_0
\; ,
\end{equation}
where the action of the different operators in the computational basis
is
\begin{align}\label{eq:hamiltoniandgonebodytermbyterm}
\nonumber
  B_i \, 
& \lvert
     g_1, \, \ldots, \, g_r; \, k_1, \, \ldots, k_s; \,
     \ell_1, \, \ell_2
  \rangle
\\
\nonumber
&=
  \delta_e ( g_i ) \,
  \lvert
     g_1, \, \ldots, \, g_r; \, k_1, \, \ldots, k_s; \,
     \ell_1, \, \ell_2
  \rangle
\; ,
\\
\nonumber
  B_0 \, 
& \lvert
     g_1, \, \ldots, \, g_r; \, k_1, \, \ldots, k_s; \,
     \ell_1, \, \ell_2
  \rangle
\\
\nonumber
&=
  \delta_e (
    g_1^{-1} \ldots g_r^{-1} \ell_2^{-1} \ell_1^{-1} \ell_2 \ell_1
  ) \,
  \lvert
     g_1, \, \ldots, \, g_r; \, k_1, \, \ldots, k_s; \,
     \ell_1, \, \ell_2
  \rangle
\; ,
\\
\nonumber
  A_j \, 
& \lvert
     g_1, \, \ldots, \, g_r; \, k_1, \, \ldots, k_s; \,
     \ell_1, \, \ell_2
  \rangle
\\
\nonumber
&=
  \frac{ 1 }{ \lvert G \rvert } \,
  \sum_{ n \in G }
  \lvert
     g_1, \, \ldots, \, g_r; \,
     k_1, \, \ldots, k_j n^{-1}, \, \ldots, \, k_s; \,
     \ell_1, \, \ell_2
  \rangle
\; ,
\\
\nonumber
  A_0 \, 
& \lvert
     g_1, \, \ldots, \, g_r; \, k_1, \, \ldots, k_s; \,
     \ell_1, \, \ell_2
  \rangle
\\
&=
  \frac{ 1 }{ \lvert G \rvert } \,
  \sum_{ n \in G }
  \lvert
     n g_1 n^{-1}, \, \ldots, \, n g_r n^{-1}; \,
     n k_1, \, \ldots, n k_s; \,
     n \ell_1 n^{-1}, \, n \ell_2 n^{-1}
  \rangle
\; .
\end{align}
Each $A_i$, $B_j$ is thus a one-body operator, and the terms $A_0$ and
$B_0$ couple all qudits (except for the electric qubits in the case of
$B_0$).  This generalises Hamiltonian
(\ref{bubbles:twobytwohamiltonian}) for the toric code on the same
lattice.  Note that the precise location of the spikes is immaterial.

\begin{figure}[htbp]
\centering
\includegraphics{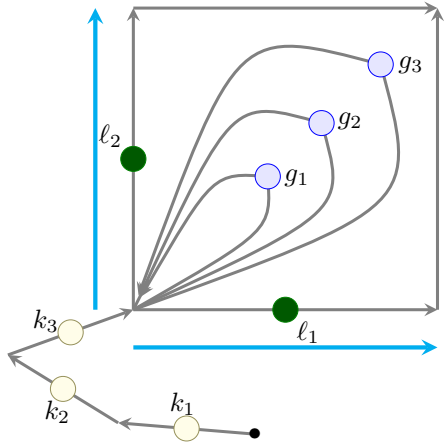}
\caption{\label{fig:dgising} {\small \textbf{Classical chains for
      quantum doubles.}  The lattice for the expression
    (\ref{eq:hamiltoniandgising}) of the Hamiltonian mostly as a sum
    of two-body terms.  The notation is as in figure
    \ref{fig:dgonebodyterms}.}}
\end{figure}
In figure \ref{fig:dgising} we give our conventions for the lattice
generalising the double-Ising-chain construction for the toric code.
The Hamiltonian is now
\begin{equation}\label{eq:hamiltoniandgising}
  H
=
 - \, B_1
 - \, \sum_{i=1}^{r-1} B_{i, \, i+1}
 - \, B_{r, \, 0}
 - \, A_1
 - \, \sum_{j=1}^{s-1} A_{j, \, j+1}
 - \, A_{s, \, 0}
\; ,
\end{equation}
where the different terms are given by
\begin{align}\label{eq:hamiltoniandgisingtermbyterm}
\nonumber
  B_1 \, 
& \lvert
     g_1, \, \ldots, \, g_r; \, k_1, \, \ldots, k_s; \,
     \ell_1, \, \ell_2
  \rangle
\\
\nonumber
&=
  \delta_e ( g_1 ) \,
  \lvert
     g_1, \, \ldots, \, g_r; \, k_1, \, \ldots, k_s; \,
     \ell_1, \, \ell_2
  \rangle
\; ,
\\
\nonumber
  B_{i, \, i+1} \, 
& \lvert
     g_1, \, \ldots, \, g_r; \, k_1, \, \ldots, k_s; \,
     \ell_1, \, \ell_2
  \rangle
\\
\nonumber
&=
  \delta_e ( g_i g_{i+1}^{-1} ) \,
  \lvert
     g_1, \, \ldots, \, g_r; \, k_1, \, \ldots, k_s; \,
     \ell_1, \, \ell_2
  \rangle
\; ,
\\
\nonumber
  B_{r, \, 0} \, 
& \lvert
     g_1, \, \ldots, \, g_r; \, k_1, \, \ldots, k_s; \,
     \ell_1, \, \ell_2
  \rangle
\\
\nonumber
&=
  \delta_e (
    g_r^{-1} \ell_2^{-1} \ell_1^{-1} \ell_2 \ell_1
  ) \,
  \lvert
     g_1, \, \ldots, \, g_r; \, k_1, \, \ldots, k_s; \,
     \ell_1, \, \ell_2
  \rangle
\; ,
\\
\nonumber
  A_1 \, 
& \lvert
     g_1, \, \ldots, \, g_r; \, k_1, \, \ldots, k_s; \,
     \ell_1, \, \ell_2
  \rangle
\\
\nonumber
&=
  \frac{ 1 }{ \lvert G \rvert } \,
  \sum_{ n \in G }
  \lvert
     g_1, \, \ldots, \, g_r; \,
     k_1 n^{-1}, \, k_2, \, \ldots, k_j, \, \ldots, \, k_s; \,
     \ell_1, \, \ell_2
  \rangle
\; ,
\\
\nonumber
  A_{ j, \, j+1 } \, 
& \lvert
     g_1, \, \ldots, \, g_r; \, k_1, \, \ldots, k_s; \,
     \ell_1, \, \ell_2
  \rangle
\\
\nonumber
&=
  \frac{ 1 }{ \lvert G \rvert } \,
  \sum_{ n \in G }
  \lvert
     g_1, \, \ldots, \, g_r; \,
     k_1, \, \ldots, n k_j, \, k_{j+1} n^{-1}, \, \ldots, \, k_s; \,
     \ell_1, \, \ell_2
  \rangle
\; ,
\\
\nonumber
  A_{ s, \, 0 } \, 
& \lvert
     g_1, \, \ldots, \, g_r; \, k_1, \, \ldots, k_s; \,
     \ell_1, \, \ell_2
  \rangle
\\
&=
  \frac{ 1 }{ \lvert G \rvert } \,
  \sum_{ n \in G }
  \lvert
     n g_1 n^{-1}, \, \ldots, \, n g_r n^{-1}; \,
     k_1, \, \ldots, , \, k_{s-1}, \, n k_s; \,
     n \ell_1 n^{-1}, \, n \ell_2 n^{-1}
  \rangle
\; .
\end{align}
Terms $B_1$, $A_1$ are one-body, and terms $B_{i, \, i+1}$ and $A_{j,
  j+1}$ are two-body (and define spin models which are essentially
classical since all these terms commute); the projector $B_{r, \, 0}$
couples one of the magnetic qudits with the topological edges (that
this is not a many-body operator is one of the simplifications with
respect to the general Hopf algebraic case); and $A_{s, \, 0}$ couples
the topological qudits with one of the electric edges and all of the
magnetic edges.  This generalises the double-Ising-chain mapping, with
the terms $B_{r, \, 0}$, $A_{s, \, 0}$ representing the toric boundary
conditions.  The topological degeneracy is recovered easily: because
of the one-body projectors $B_1$, $A_1$ breaking the degeneracy of the
two-body chains, the ground states of (\ref{eq:hamiltoniandgising})
feature product states in all the magnetic and electric qudits; and
the remaining degrees of freedom are just the topological qudits, that
is, the edges of the minimal torus studied in section
\ref{subsec:qdoubles:smallesttorus}, with the Hamiltonian $- B_P -
A_V$ defined by the operators in (\ref{eq:smalltorusavbp}).
Explicitly, the ground states of (\ref{eq:hamiltoniandgising}) are
\begin{equation}\label{eq:groundstatesdgising}
  \frac{ 1 }{ \lvert G \rvert^{(s+1)/2} } \,
  \sum_{ n_1, \, \ldots, n_s, \, n_0 \in G }
  \lvert
     e, \, \ldots, \, e; \, n_1, \, \ldots, n_s; \,
     n_0 \ell_1 n_0^{-1}, \, n_0 \ell_2 n_0^{-1}
  \rangle
\end{equation}
where $\ell_1$, $\ell_2$ are mutually commuting elements of $G$.

\subsection{\label{subsec:qdoubles:intertwining}%
  The algebraic meaning}

The significance of the elementary moves goes beyond the mere fact
that they intertwine the quantum double Hamiltonians in the two
lattices.  More fundamentally, accompanying the geometrical evolution
there is a topological charge redistribution more detailed than just
the reshuffling of binary magnetic and electric labels.

\begin{figure}[htbp]
\centering
\includegraphics{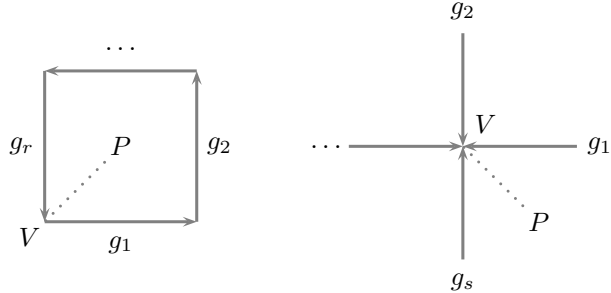}
\caption{\label{fig:conventionsrepnshdg} {\small \textbf{The quantum
      double algebra.} Conventions for the definition of plaquette and
    vertex representations (\ref{eq:defplaquetterepresentations}),
    (\ref{eq:defvertexrepresentations}).}}
\end{figure}
Consider again the problem of the local distribution of topological
charges in an arbitrary state of a quantum double model. Instead of
characterising this distribution of charges by a collection of charge
labels at different sites (recall that a site is a pair of
neighbouring plaquette and vertex, and that charge labels in general
do not split cleanly into a magnetic and electric part), we can choose
to endow the system with an algebra of local operators generalising
projectors (\ref{eq:defplaquetteprojectors}) and
(\ref{eq:defvertexprojectors}).  Plaquette operators are
representations of the algebra of functions $f$ from $\mathbb{C} G$ to
the complex numbers,
\begin{equation}\label{eq:defplaquetterepresentations}
  B_{P,V} ( f ) \,
  \lvert g_1, \, g_2, \, \ldots, \, g_r \rangle
=
  f ( g_r \ldots g_2 g_1 ) \,
  \lvert g_1, \, g_2, \, \ldots, \, g_r \rangle
\; ,
\end{equation}
and vertex operators are representations of $\mathbb{C} G$, defined by
\begin{equation}\label{eq:defvertexrepresentations}
  A_{V,P} ( k ) \,
  \lvert g_1, \, g_2, \, \ldots, \, g_s \rangle
=
  A_V ( k ) \,
  \lvert g_1, \, g_2, \, \ldots, \, g_s \rangle
=
  \lvert k g_1, \, k g_2, \, \ldots, \, k g_s \rangle
\; ,
\end{equation}
in the conventions of figure \ref{fig:conventionsrepnshdg}.  It can be
shown that the $B_{P,V} (f)$ and $A_{V,P} (k)$ operators acting on a
site define a reducible representation of $\mathrm{D} (G)$,
decomposing into the different charge sectors for that site.  Hence,
we can simply say that the charge distribution is determined
\emph{locally} by the reaction of the lattice state to operators
(\ref{eq:defplaquetterepresentations}) and
(\ref{eq:defvertexrepresentations}), for the different plaquettes and
vertices.  This has the advantage that we can work with plaquettes and
vertices separately, without the complications due to the intricate
nature of dyonic charges.  Note that vertex operators
(\ref{eq:defvertexrepresentations}) do not depend on the plaquette;
this property is lost in the more general Hopf algebraic setting of
\cite{BMCA}.

The deepest property of the elementary moves is that they are
\textsl{intertwiners} for representations $A_{V,P} (g)$ and $B_{P,V}
(f)$ along the flow of lattice deformations they define.  (Ultimately,
these moves realise morphisms in a category of representations.)  The
most complete approach to this deep property would entail a full study
of the transformations of Kitaev's ribbon operators
\cite{Kitaev:1997}, the building blocks of the representations $A_V
(g)$ and $B_P (f)$, under elementary moves, but for this paper we will
just cover the transformation of the latter, and do so using figures. 

The general form of the intertwining property is
\begin{equation}\label{eq:generalintertwining}
  U_{s_0} \, A_s (g)
=
  A_{s'} (g) \, U_{s_0} 
\; , \qquad 
  U_{s_0} \, B_s (f)
=
  B_{s'} (f) \, U_{s_0} 
\; ,
\end{equation}
where the move is either a plaquette or a vertex move associated with
site $s_0$, and $s'$ is the image of $s$ after the lattice
deformation.  This obviously implies the intertwining property for the
Hamiltonian, but is much more general.  The transformation of the
sites is subtle, though, and the notion of site as strict neighbouring
pair has to be relaxed.  Let us consider the examples that we already
used to define the general moves.

\begin{figure}[htbp]
\centering
\includegraphics{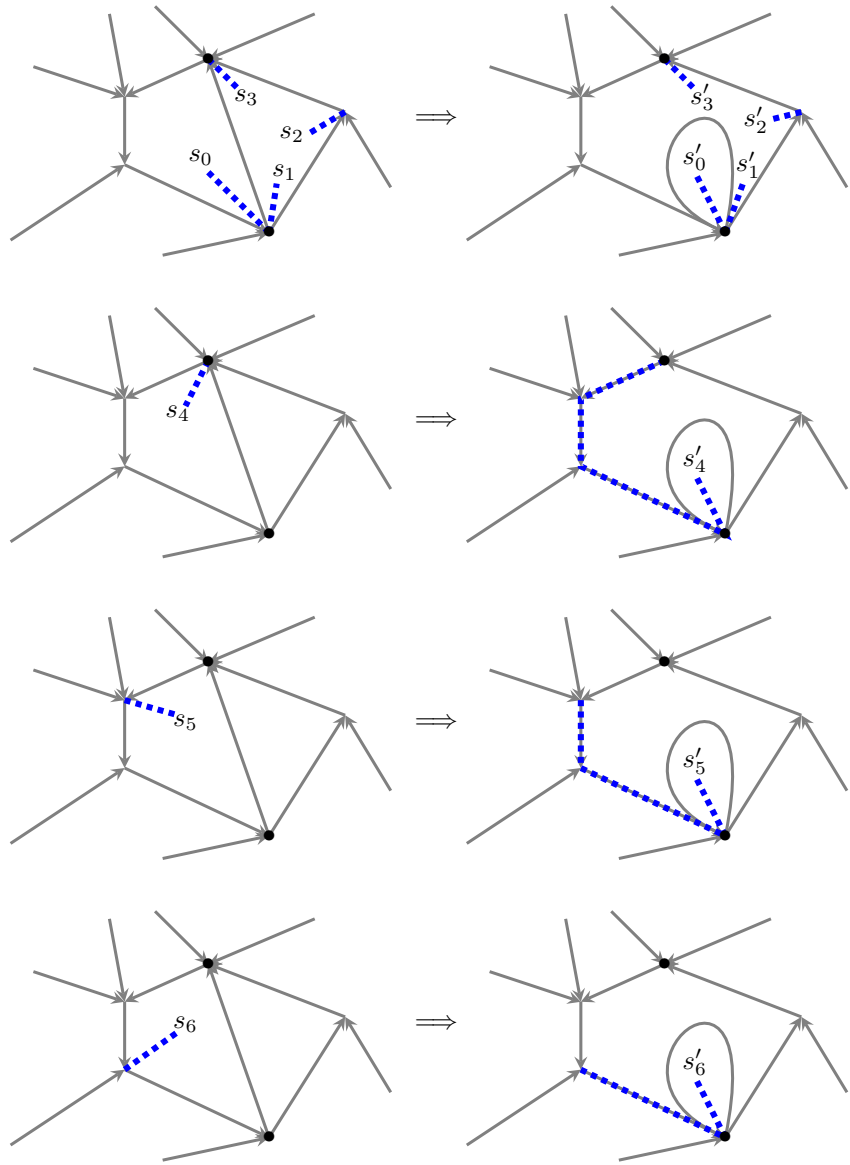}
\caption{\label{fig:dgpmovesites} {\small \textbf{Transformation of
      sites upon P-moves.}  The $P$-move of figure \ref{fig:dgpmove}
    is associated with site $s_0$.  Sites having the plaquette $P_0$
    in common with $s_0$ have a subtle transformation involving a path
    along the boundary of this plaquette.}}
\end{figure}
Any elementary move deforms one edge, which talks to two plaquettes
$P_0, \, P_1$ and two vertices $V_0, \, V_1$.  In our convention,
moves are associated with one site, say $s_0 = (P_0, \, V_0)$.
Consider the P-move studied in figure \ref{fig:dgpmove}, and the
transformation of representations $A_{V,\, P} (g)$ and $B_{P,V} (f)$
encoded in the geometric transformation of sites.  In general, sites
transform as $s = (P, \, V) \mapsto s' = (P', \, V')$ according to the
images of $P$ and $V$ under the deformation of the lattice (first row
of figure \ref{fig:dgpmovesites}).  However, sites of the form $s =
(P_0, V)$ with $V \neq V_0$ have a more subtle transformation, as
shown in the last three rows of figure \ref{fig:dgpmovesites}.  In
those cases, $s' = (P'_0{}^{V'V'_0}, V'_0)$, where $P'_0{}^{V'V'_0}$
is the plaquette $P'_0$ \textsl{conjugated} by the path from $V'$ to
$V'_0$ along the boundary of $P_0$.  Operationally, this means that
representations $A_V$ get transformed into $A_{V'}$, and $B_P$ gets
transformed into an operator whose argument is the conjugation of the
group element along the boundary edge of $P'_0$ by the product of
group elements along the path $V'V'_0$ (this conjugation is invisible
for the operators entering the Hamiltonian).

\begin{figure}[htbp]
\centering
\includegraphics{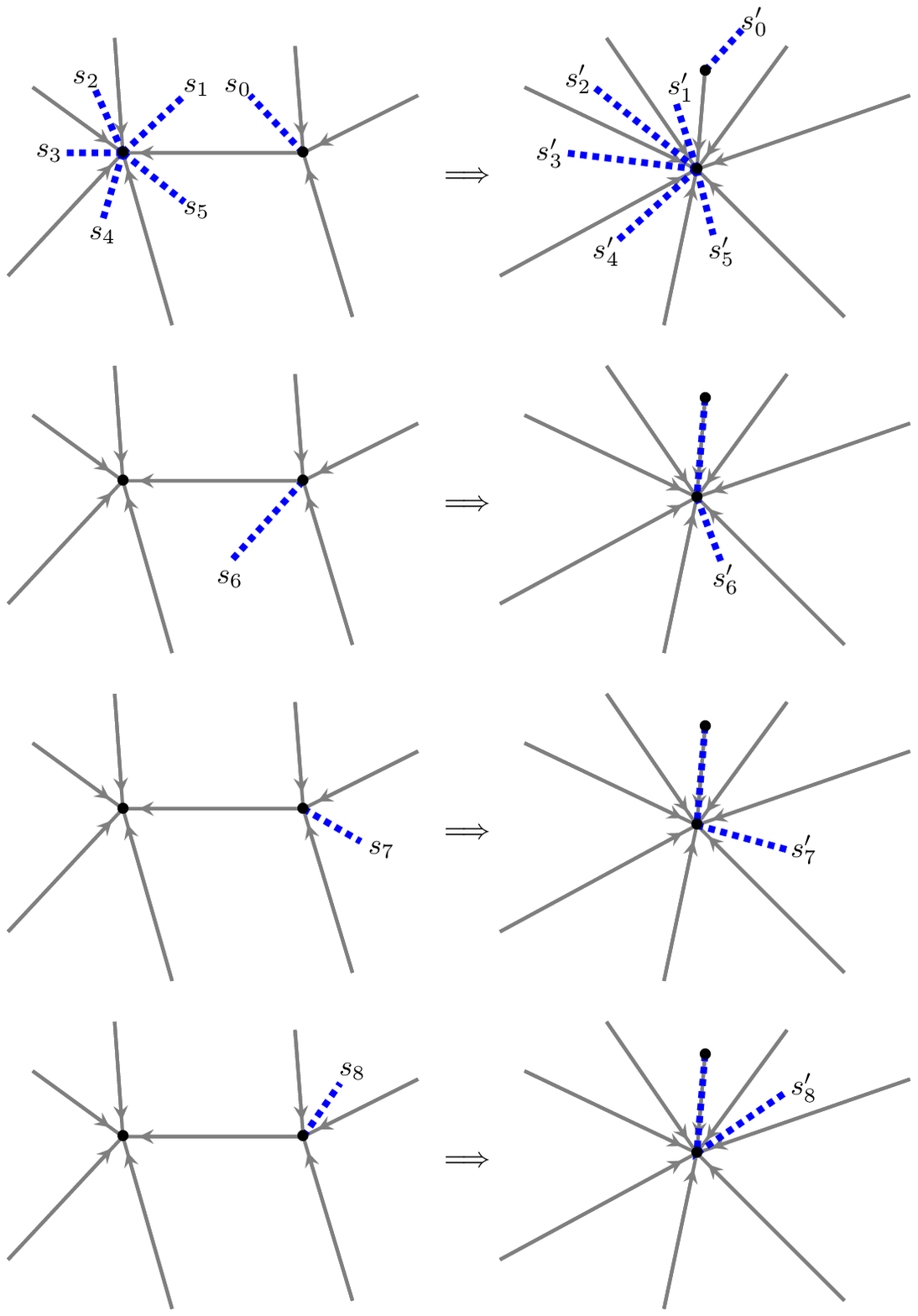}
\caption{\label{fig:dgvmovesites} {\small \textbf{Transformation of
      sites under V-moves.}  The move in figure \ref{fig:dgvmove} is
    associated with site $s_0$.  Sites having the vertex in common
    with $s_0$ have again a nontrivial transformation.}}
\end{figure}
V-moves can be analysed in much the same way.  Consider the move in
figure \ref{fig:dgvmove}, associated with site $s_0 = (P_0, \, V_0)$.
Sites transform again as $s = (P, \, V) \mapsto s' = (P', \, V')$
except for those having $V_0$ as a vertex and $P \neq P_0$ (note that
$s_0$ itself transforms into $(P'_0, V'_0)$).  We have $s = (P, \,
V_0) \mapsto (P'{}^{V'_0V'_1}, V'_0)$, that is, the plaquette is
conjugated by a path from $V'_0$ to $V'_1$ (see figure
\ref{fig:dgvmovesites}).

\clearpage

\section{\label{sec:discussion}%
  Discussion}

In summary, we have extended the entanglement renormalisation approach
for ground states of topological lattice systems to a unitary quantum
circuit disentangling procedure that diagonalises the Hamiltonian.  We
give this construction for Kitaev's quantum double models based on
groups, on the understanding that the construction generalises
directly to more general quantum double models based on Hopf
$C^\ast$-algebras.  The extension of the Abelian construction to the
non-Abelian case is informed by the understanding of the structure of
the Hilbert space and energy levels in the toric topology, and we give
an account of this with details and a specific example to be found in
the appendices.

While this ER picture extended to the full spectrum of the quantum
doubles is geometrically pleasant, it is probably not a practical way
to construct the relevant codes from a small instance by adding
degrees of freedom, since it requires very nonlocal operations.

Note that the double Ising chain lurking behind the toric code has
been exposed by different authors.  In particular, a duality mapping
was given in \cite{NussinovOrtiz} showing that the partition function
of the toric code is identical to that of two noninteracting Ising
chains.  The construction presented here is graphic and explicit, and
lends itself to generalisation for non-Abelian quantum double models.
It also uncovers an intriguing geometrical picture of renormalisation,
proceeding through deformations simplifying the original lattice, that
can probably be extended to general string-net models building upon
the work in \cite{Koenig}.

On the other hand, what we achieve in the end is (up to boundary
conditions) the diagonalisation of a Hamiltonian consisting of
mutually commuting projectors.  The approach of
Ref.~\cite{Latorre:circuit} is a general method for diagonalising
strongly correlated Hamiltonians in terms of a quantum circuit, and
the present work can be regarded as an illustration thereof (the
tensor network behind this construction being a unitary quantum
circuit obtained from unitarisation of the ER scheme of
\cite{AguadoVidal}).

As remarked in the introduction, a lattice is essentially a
discretisation of the metric in a continuum theory.  One of the
notable points of this construction is that the flow of the quantum
circuit is universal for all the quantum double models, being encoded
in the simplifying moves acting on the lattice structure: whether this
points to a meaningful procedure in the continuum limit is a
suggestive question.

Work using related geometrical ideas, leading to numerical methods for
the $\mathbb{Z}_2$ gauge theory, was reported in \cite{Luca}.  Similar
ideas are being used by R.~K\"onig, F.~Verstraete, and collaborators,
working towards a more general renormalisation scheme for lattice
systems, to be found in \cite{KoenigVerstraete}.

\subsubsection*{\label{acks}%
  Acknowledgements}

I thank Guifr\'e Vidal, Gavin Brennen, Oliver Buerschaper and Sofyan
Iblisdir for key discussions on quantum double models, Robert K\"onig
and Frank Verstraete for generous communication about their work, and
the Kavli Institute for Theoretical Physics for hospitality during the
research programme \textsl{Disentangling quantum many-body systems:
  Computational and conceptual approaches}.  This research was
supported in part by the National Science Foundation under Grant
No.~NSF PHY05-51164.

\clearpage

\appendix

\section{\label{sec:appendix:dgmodels}%
  Topological charges in $\mathrm{D} (G)$ models}

In a $\mathrm{D} (G)$ model, topological charges are given in terms of
the group structure and its representation theory (in the recent paper
\cite{Beigi}, it was shown that not all such charges are physically
distinguishable; however, we will not be concerned with these issues
here).  Charge labels are pairs $( C, \, \alpha )$, where:
\begin{itemize}
\item %
  $C$ runs over conjugacy classes of the group, that is, orbits of
  group elements under the adjoint action $t \mapsto g t g^{-1}$.
\item %
  $\alpha$ runs over irreducible representations (irreps) of the
  \textsl{centraliser} group $N_C$ of the conjugacy class $C$.
\end{itemize}

The centraliser $N_C$ is constructed as follows: given $c \in C$, its
centraliser $N_c$ is the subgroup of $G$ consisting of all the
elements of $G$ commuting with $c$,
\begin{equation}\label{eq:defcentralisercsmall}
  N_c = \{ n \in G; \; n c n^{-1} = c \} \; .
\end{equation}
It is immediately checked that
\begin{equation}\label{eq:conjugationcentralisers}
  n \in N_c
\quad \Rightarrow \quad
  g n g^{-1} \in N_{ g c g^{-1} }
\; ,
\end{equation}
so the centralisers of elements of the same conjugacy class are
conjugate (and hence, isomorphic) to each other, justifying the
definition of the abstract $N_C$.  The size of this group is $\lvert
N_C \rvert = \lvert G \rvert / \lvert C \rvert$.

Pairs $( C, \, \alpha )$ correspond to irreducible representations of
the quantum double $\mathrm{D} (G)$; for the purposes of this work, it
suffices to present $\mathrm{D} (G)$ as the tensor product
$\mathbb{C}^G \otimes \mathbb{C} G$ of the space of complex functions
on the group and the space of complex linear combinations of group
elements, or group algebra.  This vector space is spanned by elements
$P_h g \equiv \delta_h \otimes g$, where $g, \, h \in G$, and
$\delta_h : k \mapsto \delta_{ h, \, k }$ are Kronecker delta
functions.  The algebra structure in $\mathrm{D} (G)$ is given by the
multiplication
\begin{equation}\label{eq:multdg}
  P_h g \cdot P_k \ell
=
  \delta_h ( g k g^{-1} ) \, P_h g \ell
\; .
\end{equation}

An irreducible representation $( C, \, \alpha )$ of $\mathrm{D} (G)$
realises elements $P_h g$ as endomorphisms of a linear representation
space $V^{ ( C, \, \alpha ) }$; this can be identified with
$\mathbb{C} C \otimes V^\alpha$, where $V^\alpha$ is the
representation space for group representation $\alpha$.  A basis of
$V^{ ( C, \, \alpha ) }$ is given by $\{ \lvert k; \, v_i \rangle \}$,
with $k \in C$ and $\{ \lvert v_i \rangle \}$ a basis of $V^\alpha$,
such that
\begin{equation}\label{eq:irrepalpha}
  D^\alpha ( n ) \,
  \lvert v_i \rangle
=
  \sum_j
  \lvert v_j \rangle \,
  D^\alpha_{ji} ( n )
\; , \quad
  n \in N_C \; .
\end{equation}
The matrix form of $( C, \, \alpha )$ reads \cite{Gould, Overbosch}
\begin{equation}\label{eq:matrixformcalpha}
  D^{ ( C, \, \alpha ) } ( P_h g ) \,
  \lvert k, \, v_i \rangle
=
  \delta_h ( g k g^{-1} ) \,
  \sum_j
  \lvert g k g^{-1}, \, v_j \rangle \,
  D^\alpha_{ji} ( g \rvert_{N_{k_0}} )
\; ,
\end{equation}
or, in other words,
\begin{equation}\label{eq:matrixformcalphabis}
  D^{ ( C, \, \alpha ) }_{ \ell j, \, k i } ( P_h g )
=
  \delta_\ell ( g k g^{-1} ) \,
  \delta_h    ( g k g^{-1} ) \,
  D^\alpha_{ji} ( g \rvert_{N_{k_0}} )
\; .
\end{equation}
Here $k_0$ is an arbitrarily chosen reference element in conjugacy
class $C$, and $\alpha$ is chosen as a representation of the
normaliser $N_{k_0}$. The notation $g \rvert_{N_{k_0}}$ means a
projection of $g \in G$ onto $N_{k_0}$, defined as follows: given the
reference $k_0 \in C$, define a system of group elements $x_{k_0}^k
\in G$, one for each $k \in C$, such that $x_{k_0}^k k_0 ( x_{k_0}^k
)^{-1} = k$, and in particular $x_{k_0}^{k_0} = e$.  It can be checked
that $x_{k_0}^k N_{k_0}$ are mutually disjoint cosets whose union is
$G$.  Now the adjoint action of $g$ maps $k$ to $g k g^{-1}$, and from
this follows that $g \rvert_{N_{k_0}} \equiv ( x_{k_0}^{gkg^{-1}}
)^{-1} g x_{k_0}^k$ is an element of $N_{k_0}$, which is what appears
in (\ref{eq:matrixformcalpha}) and (\ref{eq:matrixformcalphabis}).
Different choices of $k_0$ and $\{ x_{k_0}^k \}$ give rise to
equivalent representations.

The left regular representation, that is, the representation of
$\mathrm{D} (G)$ on $\mathrm{D} (G)$ defined by left multiplication
\begin{equation}\label{eq:leftregrep}
  a \mapsto L ( a )
\; , \quad
  L ( a ) : b \mapsto a b
\; ,
\end{equation}
is reducible and contains each one of the irreps $( C, \, \alpha )$ of
$\mathrm{D} (G)$ with a multiplicity equal to its dimension $\lvert C
\rvert \lvert \alpha \rvert$, just as for finite groups:
\begin{equation}\label{eq:decompleftregular}
  \mathrm{D} (G)
\approx
  \bigoplus_{ ( C, \, \alpha ) }
  \bigoplus_{a=1}^{ \lvert C \rvert \lvert \alpha \rvert }
  V^{ ( C, \, \alpha ) }_a
\; .
\end{equation}

Explicitly, the following elements of $\mathrm{D} (G)$:
\begin{equation}\label{eq:irrepsdgwithinregular}
  [ ( C, \, \alpha ) \, k, \, \ell ; \, i, \, s ]
=
  \sqrt{
    \frac{
       \lvert C \rvert \lvert \alpha \rvert
    }{
       \lvert G \rvert
    }
  } \,
  P_k \,
  \sum_{ n \in N_k }
  D^\alpha_{si}
    \big( ( x_{k_0}^k )^{-1} n^{-1} x_{k_0}^k \big) \,
  n \, x_{k_0}^k \, ( x_{k_0}^\ell )^{-1}
\, , 
\end{equation}
where $k, \, m \in C$, are a basis of the quantum double algebra
satisfying
\begin{align}\label{eq:irrepsdgwithinregularinaction}
\nonumber
  P_h g
& \, \cdot \,
  [ ( C, \, \alpha ) \, k, \, \ell ; \, i, \, s ]
\\
&=
  \delta_h ( g k g^{-1} ) \,
  \sum_j
  [ ( C, \, \alpha ) \, g k g^{-1}, \, \ell ; \, j, \, s ] \;
  D^\alpha_{ji}
    \big( ( x_{k_0}^{ g k g^{-1} } )^{-1} g x_{k_0}^k \big)
\, , 
\end{align}
and hence, they realise decomposition (\ref{eq:decompleftregular}) if
we identify index $a$ with the pair $( \ell, \, s )$.

Projection onto the different $( C, \, \alpha )$ sectors in
$\mathrm{D} (G)$ is achieved by multiplication with the following
central algebra elements:
\begin{equation}\label{eltprojectorsontocalpha}
  \mathcal{Q}^{( C, \, \alpha )}
=
  \frac{
    \lvert C \rvert \lvert \alpha \rvert
  }{
    \lvert G \rvert
  } \,
  \sum_{ k \in C }
  \sum_{ n \in N_k }
  \chi^\alpha \big( ( x_{k_0}^k )^{-1} n^{-1} x_{k_0}^k \big) \,
  P_k \, n
\; ,
\end{equation}
where $\chi^\alpha$ is the character of representation $\alpha$.  It
can be checked that they form a decomposition of the unit into a sum
of projectors:
\begin{equation}\label{eltprojectorssums}
  \mathcal{Q}^{( C, \, \alpha )}
 \cdot
  \mathcal{Q}^{( C', \, \alpha' )}
=
  \delta^{ ( C, \, \alpha ), \, ( C', \, \alpha' ) } \,
  \mathcal{Q}^{( C, \, \alpha )}
\; , \quad
  \sum_{ ( C, \, \alpha ) }
  \mathcal{Q}^{( C, \, \alpha )}
=
  1_{ \mathrm{D} (G) }
=
  \sum_h P_h e
\; .
\end{equation}
Upon multiplication, they project onto the irrep blocks $\bigoplus_a
V^{ ( C, \, \alpha ) }_a$ of decomposition
(\ref{eq:decompleftregular}):
\begin{equation}\label{eltprojectorsontocalphaproperty} 
  \mathcal{Q}^{( C, \, \alpha )} 
 \, \cdot \,
  [ ( C', \, \alpha' ) \, k, \, \ell ; \, i, \, s ]
=
  \delta^{ ( C, \, \alpha ), \, ( C', \, \alpha' ) } \,
  [ ( C', \, \alpha' ) \, k, \, \ell ; \, i, \, s ]
\; .
\end{equation}

The key algebraic structure defined on the $\mathrm{D} (G)$ lattice
models is the algebra of \textsl{ribbon operators} \cite{Kitaev:1997}
(see also \cite{BombinMD} for a very readable introduction).  Without
going into the details of the definition, these operators are
assembled from one-edge operators $L_\pm ( g )$ representing $G$ (or,
more generally, the group algebra $\mathbb{C} G$), and $T_\pm ( f )$
representing the dual algebra $\mathbb{C}^G$ of complex functions on
$G$:
\begin{equation}\label{eq:eltribbons}
  L_+ ( g ) \, a = g a
\, , \quad
  L_- ( g ) \, a = a g^{-1}
\, , \quad
  T_+ ( f ) \, a = f (a) \, a
\, , \quad
  T_- ( f ) \, a = f ( a^{-1} ) \, a
\; .
\end{equation}
The $L_\pm$ are just the left and right regular representations of
$\mathbb{C} G$, while the $T_\pm$ are equivalent to the left and right
regular representations of $\mathbb{C}^G$ (through the map $g \mapsto
\sqrt{ \lvert G \rvert } \, \delta_{ g^{-1} }$, which is an
electric-magnetic duality \cite{BCKA}).

By forming combinations of these elementary operations using the
properties of $\mathbb{C} G$ and $\mathbb{C}^G$ as Hopf algebras, one
can build representations of $\mathrm{D} (G)$ and its dual.  In
particular, one can define projectors to `measure the topological
charge' associated with geometrical closed ribbons by representing the
elements $\mathcal{Q}^{ ( C, \, \alpha ) } \in \mathrm{D} (G)$.

Coming back to our smallest torus, we define topological labels for a
horizontal loop, a vertical loop, and the bulk site, using the
following representations of $\mathrm{D} (G)$:
\begin{align}\label{eq:ribbonrepresentations}
\nonumber
  F_{ \mathrm{hor} } ( P_h g ) \,
  \lvert a, \, b \rangle_{ \mathrm{cb} }
&=
  \delta_h ( g a g^{-1} ) \,
  \lvert g a g^{-1}, \, g b \rangle_{ \mathrm{cb} }
\; ,
\\
\nonumber
  F_{ \mathrm{vert} } ( P_h g ) \,
  \lvert a, \, b \rangle_{ \mathrm{cb} }
&=
  \delta_{ h^{-1} } ( g b g^{-1} ) \,
  \lvert g a, \, g b g^{-1} \rangle_{ \mathrm{cb} }
\; ,
\\
  F_{ \mathrm{bulk} } ( P_h g ) \,
  \lvert a, \, b \rangle_{ \mathrm{cb} }
&=
  \delta_h ( g b^{-1} a^{-1} b a g^{-1} ) \,
  \lvert g a g^{-1}, \, g b g^{-1} \rangle_{ \mathrm{cb} }
\; .
\end{align}
{}From these representations, the projectors onto definite topological
sectors are
\begin{align}\label{eq:ribbonprojectors}
\nonumber
  F_{ \mathrm{hor} }
  \big( \mathcal{Q}^{ ( C, \, \alpha ) } \big) \,
  \lvert a, \, b \rangle_{ \mathrm{cb} }
&=
  \frac{
    \lvert C \rvert \lvert \alpha \rvert
  }{
    \lvert G \rvert
  } \,
  \sum_{ k \in C }
  \delta_k ( a ) \,
  \sum_{ n \in N_k }
  \chi^\alpha \big( ( x_{k_0}^k )^{-1} n^{-1} x_{k_0}^k \big) \,
  \lvert
    a, \, n b 
  \rangle_{ \mathrm{cb} }
\; ,
\\
\nonumber
  F_{ \mathrm{vert} }
  \big( \mathcal{Q}^{ ( C, \, \alpha ) } \big) \,
  \lvert a, \, b \rangle_{ \mathrm{cb} }
&=
  \frac{
    \lvert C \rvert \lvert \alpha \rvert
  }{
    \lvert G \rvert
  } \,
  \sum_{ k \in C }
  \delta_{ k^{-1} } ( b ) \,
  \sum_{ n \in N_k }
  \chi^\alpha \big( ( x_{k_0}^k )^{-1} n^{-1} x_{k_0}^k \big) \,
  \lvert
    n a, \, b
  \rangle_{ \mathrm{cb} }
\; ,
\\
\nonumber
  F_{ \mathrm{bulk} }
  \big( \mathcal{Q}^{ ( C, \, \alpha ) } \big) \,
  \lvert a, \, b \rangle_{ \mathrm{cb} }
&
\\
=
  \frac{
    \lvert C \rvert \lvert \alpha \rvert
  }{
    \lvert G \rvert
  } \,
  \sum_{ k \in C }
& \delta_k ( b^{-1} a^{-1} b a ) \,
  \sum_{ n \in N_k }
  \chi^\alpha \big( ( x_{k_0}^k )^{-1} n^{-1} x_{k_0}^k \big) \,
  \lvert
    n a n^{-1}, \, n b n^{-1}
  \rangle_{ \mathrm{cb} }
\; .
\end{align}

Projectors $F_{ \mathrm{bulk} } \big( \mathcal{Q}^{ ( C, \, \alpha ) }
\big)$ commute both with horizontal and with vertical projectors in
(\ref{eq:ribbonprojectors}).  This justifies a labelling of states in
the small torus Hilbert space by a bulk charge and a loop charge
(horizontal or vertical).  The bulk charge is relevant for the energy,
since the Hamiltonian penalises the departure from trivial charge in
the bulk; loop labels, on the other hand, do not play a r\^ole for the
Hamiltonian.

The ground level of the quantum double Hamiltonian is the $+1$
eigenspace of the projector
\begin{equation}\label{eq:bulkvacuumprojector}
  F_{ \mathrm{bulk} }
  \big( \mathcal{Q}^{ ( C_e, \, 1_+ ) } \big) \,
  \lvert a, \, b \rangle_{ \mathrm{cb} }
=
  \frac{ 1 }{ \lvert G \rvert } \,
  \sum_{ n \in G }
  \delta_e ( b^{-1} a^{-1} b a ) \,
  \lvert
    n a n^{-1}, \, n b n^{-1}
  \rangle_{ \mathrm{cb} }
\; .
\end{equation}
It is easy to check that the dimension of this vacuum subspace equals
the number of topological charge labels.  We can give an orthonormal
basis labelled by a charge label in the horizontal loop, i.~e., the
$\mathrm{h}$-basis, as
\begin{equation}\label{eq:basisvacuumhorizontal}
  \lvert ( C, \, \alpha ) \rangle_{ \mathrm{h} }
=
  \frac{ 1 }{ \sqrt{ \lvert G \rvert } } \,
  \sum_{ k \in C }
  \sum_{ n \in N_k }
  \chi^\alpha \big( ( x_{k_0}^k )^{-1} n^{-1} x_{k_0}^k \big) \,
  \lvert
    k, \, n
  \rangle_{ \mathrm{cb} }
  \; ,
\end{equation}
or the alternative $\mathrm{v}$-basis labelled by vertical charges,
\begin{equation}\label{eq:basisvacuumvertical}
  \lvert ( C, \, \alpha ) \rangle_{ \mathrm{v} }
=
  \frac{ 1 }{ \sqrt{ \lvert G \rvert } } \,
  \sum_{ k \in C }
  \sum_{ n \in N_k }
  \chi^\alpha \big( ( x_{k_0}^k )^{-1} n^{-1} x_{k_0}^k \big) \,
  \lvert
    n, \, k^{-1}
  \rangle_{ \mathrm{cb} }
  \; .
\end{equation}

The overlap of basis states of (\ref{eq:basisvacuumhorizontal}) and
(\ref{eq:basisvacuumvertical}) gives the topological $S$-matrix of the
theory:
\begin{align}\label{eq:smatrixoverlap}
\nonumber
& S_{ ( C', \, \alpha' ) , \, ( C, \, \alpha ) }
=
  {}_{ \mathrm{v} } \langle ( C', \, \alpha' )
  \vert ( C, \, \alpha ) \rangle_{ \mathrm{h} }
\\
&=
  \frac{ 1 }{ \lvert G \rvert } \,
  \sum_{ k \in C }
  \sum_{ \ell \in C' \cap N_k }
  \chi^{ \alpha' }
    \big( ( x_{\ell_0}^\ell )^{-1} \, k \, x_{\ell_0}^\ell \big) \,
  \chi^\alpha \big( ( x_{k_0}^k )^{-1} \, \ell \, x_{k_0}^k \big)
\; ,
\end{align}
where $k_0, \, \ell_0$ are the reference elements of conjugacy classes
$C$, $C'$, respectively.

For the whole Hilbert space of the small torus one may expect to find
a basis labelled by a loop charge, say the horizontal label, and a
bulk charge (plus possibly internal degrees of freedom for the
latter).  This is a good strategy, but in general not all such charge
pairs are compatible.  For Abelian $G$ only the trivial charge may
appear in the bulk.  For non-Abelian models, some charges can never
appear in the bulk, and additionally some bulk charges are excluded
depending on the loop charge label.  Exactly which charges may appear
in the bulk is a property of the anyon model and ultimately a
representation-theoretical problem.  As argued in the text, the
compatible pairs of bulk charge and loop charge are those that can
fuse into the loop charge.  If the group $G$ is Abelian, the
representation theory of $\mathrm{D} (G)$ is Abelian and charges are
only invariant if they fuse with the vacuum charge $( C_e, \, 1_+ )$;
this singles out the vacuum label as the only possible bulk
charge\footnote{More consequences can be drawn; for instance, since
  the magnetic part of the bulk charge has to be a conjugacy class of
  the commutator subgroup $G' = \langle b^{-1} a^{-1} b a; \, a, \, b
  \in G \rangle$, any charge not satisfying that condition never
  leaves charges invariant upon fusion; but it is not the point of
  this paper to develop the theory of fusion further.}. For
non-Abelian models, the allowed pairs must be computed from the fusion
rules.  We give the full details for $G = \mathrm{S}_3$ in appendix
\ref{sec:appendix:dsthree}.

\begin{figure}[htbp]
\centering
\includegraphics{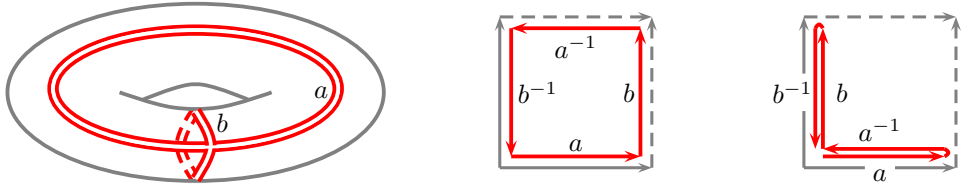}
\caption{\label{fig:bulkcharges} {\small \textbf{Definition of bulk
      charges on the smallest torus.}  Cutting a torus into a square
    defines a contractible region, with a boundary along the circuit
    $aba^{-1}b^{-1}$.  The charge measured in this region is what we
    call bulk charge.  The question of the total charge on the surface
    of the torus corresponds to the circuit $aa^{-1}bb^{-1}$, yielding
    always the trivial charge.}}
\end{figure}

A comment on the notion of charge in the bulk is in order.  The
projectors $F_{ \mathrm{bulk} } ( \mathcal{Q}^{ ( C, \, \alpha ) } )$
probe the topological charge in a contractible region, namely the
square defined by the edges (square $aba^{-1}b^{-1}$ in figure
\ref{fig:bulkcharges}); this happens to cover the whole area of the
torus but cannot see topological quantities, because it does not
include the boundary conditions.  If one measures nontrivial
topological charge on the bulk, it is exactly compensated by an
opposite charge sitting at the boundary, and measured along the
opposite circuit ($bab^{-1}a^{-1}$ in the notation of figure
\ref{fig:bulkcharges}).  All in all, the total charge on the whole
torus corresponds to a measurement along a circuit equivalent to that
in the right part of figure \ref{fig:bulkcharges}, that is, a loop
always homotopic to a point, which always yields the trivial charge.

The way in which bulk charge appears in the non-Abelian $\mathrm{D}
(G)$ models is one aspect of Kitaev's construction which differs from
a theory of gauge fields.  In a discrete gauge theory, edge variables
are gauge variant; therefore, toric boundary conditions only fix the
periodicity of the edge variables up to a gauge transformation defined
on the boundary of the bulk, which in turn is related to the bulk
charge (in, for instance, a $U (1)$ gauge theory, single-valuedness of
the gauge transformation imposes Dirac's quantisation condition, which
constrains the value of the bulk magnetic flux).  In Kitaev's lattice
models, however, edge variables correspond to physical states of
physical qudits, which are strictly periodic on the torus.

\clearpage

\section{\label{sec:appendix:dsthree}%
  $\mathrm{S}_3$ and the $\mathrm{D} ( \mathrm{S}_3 )$ model}

In this appendix we illustrate the general formalism for the smallest
non-Abelian group and the associated quantum double model.

Let $G = \mathrm{S}_3$ be the group of permutations of three objects.
It has six elements, organised into three conjugacy classes: the unit
$e$; the transpositions $t_0 = (12)$, $t_1 = (23)$, $t_2 = (31)$; and
the $3$-cycles $c_+ = (123)$, $c_- = (132)$.

The topological sectors of the $\mathrm{D} ( \mathrm{S}_3 )$ model,
that is, the irreducible representations of the quantum double algebra
$\mathrm{D} ( \mathrm{S}_3 )$, are labelled by pairs $( C, \, \alpha
)$, where $C$ is one of the conjugacy classes $C_e = \{ e \}$, $C_t =
\{ t_0, \, t_1, \, t_2 \}$, and $C_c = \{ c_+, \, c_- \}$, and
$\alpha$ is an irrep of the centraliser $N_C$ of the conjugacy class
$C$.  Since the centralisers of the three conjugacy classes of
$\mathrm{S}_3$ (shown in table~\ref{table:irrepssthree}) are $N_{ C_e
} = \mathrm{S}_3$ (with irreps $1_+$, $1_-$, $2$), $N_{ C_t } =
\mathbb{Z}_2$ (with irreps $1_+$, $1_-$), and $N_{ C_c } =
\mathbb{Z}_3$ (with irreps $1_+$, $1_\omega$, $1_{\bar \omega}$), we
have eight charges.
\begin{table}[htbp]
\centering
\begin{tabular}{c||c|c|c|c|c|c|}
  $N_{C_e} = \mathrm{S}_3$
    & $e$ & $t_0$ & $t_1$ & $t_2$ & $c_+$ & $c_-$
\\
\hline\hline
  $1_+$ & $1$ &  $1$ &  $1$ &  $1$ & $1$ & $1$ 
\\
\hline
  $1_-$ & $1$ & $-1$ & $-1$ & $-1$ & $1$ & $1$ 
\\
\hline
  $2$
    & $\begin{pmatrix} 1 & 0 \\ 0 & 1 \end{pmatrix}$
    & $\begin{pmatrix} 0 & 1 \\ 1 & 0 \end{pmatrix}$
    & $\begin{pmatrix} 0 & \omega \\ \bar\omega & 0 \end{pmatrix}$
    & $\begin{pmatrix} 0 & \bar\omega \\ \omega & 0 \end{pmatrix}$
    & $\begin{pmatrix} \bar\omega & 0 \\ 0 & \omega \end{pmatrix}$
    & $\begin{pmatrix} \omega & 0 \\ 0 & \bar\omega \end{pmatrix}$
\\
\hline
\\
\end{tabular}
\begin{tabular}{c||c|c|}
  $N_{C_t} = \mathbb{Z}_2$ & $e$ & $a$ \\
\hline\hline
  $1_+$ & $1$ &  $1$ \\
\hline
  $1_-$ & $1$ & $-1$ \\
\hline
\end{tabular}
\hspace{1cm}
\begin{tabular}{c||c|c|c|}
  $N_{C_c} = \mathbb{Z}_3$ & $e$ & $b$ & $b^2$ \\
\hline\hline
  $1_+$            & $1$ & $1$          & $1$ \\
\hline
  $1_\omega$       & $1$ & $\omega$     & $\bar\omega$ \\
\hline
  $1_{\bar\omega}$ & $1$ & $\bar\omega$ & $\omega$ \\
\hline
\end{tabular}
\caption{\label{table:irrepssthree} Irreducible representations of the
  centralisers $\mathrm{S}_3$, $\mathbb{Z}_2$, and $\mathbb{Z}_3$ of
  the conjugacy classes of $\mathrm{S}_3$.  Here $\omega =
  \mathrm{e}^{ i \, 2 \pi / 3}$.}
\end{table}

The fusion rules of these representations (i.~e., the Clebsch-Gordan
decompositions of their tensor products) are best presented in the
manner of Overbosch \cite{Overbosch}.  Labelling irreps by their
dimension, we write $1 = ( C_e, \, 1_+ )$; ${\bar 1} = ( C_e, \, 1_-
)$; $2^{x,y,z,w} = ( C_e, \, 2 )$, $( C_c, \, 1_{+,\omega,\bar\omega}
)$; and $3^{a,b} = ( C_t, \, 1_\pm )$.  Then the nontrivial fusion
rules can be written as
\begin{align}\label{eq:fusionirrepsdsthree}
\nonumber
  {\bar 1} \otimes {\bar 1} \rightarrow 1,
\qquad
  {\bar 1} \otimes 2^x &\rightarrow 2^x,
\qquad
  {\bar 1} \otimes 3^a \rightarrow 3^b,
\\
\nonumber
  2^x \otimes 2^x \rightarrow 1 \oplus {\bar 1} \oplus 2^x,
\qquad
  2^x \otimes 2^y &\rightarrow 2^z \oplus 2^w,
\qquad
  2^x \otimes 3^a \rightarrow 3^a \oplus 3^b,
\\
  3^a \otimes 3^a
\rightarrow
  1 \oplus 2^x \oplus 2^y \oplus 2^z \oplus 2^w,
&\qquad
  3^a \otimes 3^b
\rightarrow
  {\bar 1} \oplus 2^x \oplus 2^y \oplus 2^z \oplus 2^w,
\end{align}
and the rules obtained by permutations in $a, \, b$ and in $x, \, y,
\, z, \, w$.  In particular, each charge is its own conjugate.

In the following we use the shorthand $(e \pm) = ( C_e, \, 1_\pm )$;
$(e 2) = ( C_e, \, 2 )$; $(t \pm) = ( C_t, \, 1_\pm )$, and $( c \eta
) = ( C_c, \, 1_\eta )$ for the irreps of $\mathrm{D} ( \mathrm{S_3}
)$, where $\eta$ is one of the cubic roots of unity $1, \, \omega, \,
\bar\omega$.

For the explicit construction of these irreps we choose reference
elements $e \in C_e$, $t_0 \in C_t$, and $c_+ \in C_c$.  The
nontrivial fixed elements $x_{k_0}^k$ such that $x_{k_0}^k k_0 (
x_{k_0}^k )^{-1} = k$ are chosen as $x_{t_0}^{t_1} = c_+$,
$x_{t_0}^{t_2} = c_-$, $x_{c_+}^{c_-} = t_0$.  Projectors
(\ref{eltprojectorsontocalpha}) onto irrep blocks of the left regular
representation read
\begin{align}\label{eq:eltprojectorssthree}
\nonumber
  \mathcal{Q}^{ (e \pm) }
&=
  \frac{1}{6} \,
  P_e \, ( e \pm t_0 \pm t_1 \pm t_2 + c_+ + c_- )
\; ,
\\
\nonumber
  \mathcal{Q}^{ (e 2) }
&=
  \frac{1}{3} \,
  P_e \, ( 2 e - c_+ - c_- )
\; ,
\\
\nonumber
  \mathcal{Q}^{ (t \pm) }
&=
  \frac{1}{2} \,
  \big\{
     P_{t_0} \, ( e \pm t_0 )
   + P_{t_1} \, ( e \pm t_1 )
   + P_{t_2} \, ( e \pm t_2 )
  \big\}
\; ,
\\
  \mathcal{Q}^{ (c \eta) }
&=
  \frac{1}{3} \,
  \big\{
     P_{c_+} \, ( e + \bar\eta c_+ + \eta c_- )
   + P_{c_-} \, ( e + \eta c_+ + \bar\eta c_- )
  \big\}
\; .
\end{align}

In the quantum double model defined on the smallest torus of figure
\ref{fig:smalltorusdgbasis}, there are eight simultaneous $+1$
eigenstates of $A_V$ and $B_P$ (or just $+1$ eigenstates of the
projector onto vacuum bulk charge, $A_V B_P = F_{\mathrm{bulk}} (
\mathcal{Q}^{(e +)} )$):
\begin{align}\label{eq:groundstatessmalltorussthree}
\nonumber
  \lvert e, \, e \rangle_{\mathrm{cb}} ,
\quad
  \frac{ 1 }{ \sqrt{3} } \,
  \big\{
     \lvert e, \, t_0 \rangle_{\mathrm{cb}}
&  + \lvert e, \, t_1 \rangle_{\mathrm{cb}}
   + \lvert e, \, t_2 \rangle_{\mathrm{cb}}
  \big\} ,
\quad
  \frac{ 1 }{ \sqrt{2} } \,
  \big\{
     \lvert e, \, c_+ \rangle_{\mathrm{cb}}
   + \lvert e, \, c_- \rangle_{\mathrm{cb}}
  \big\} ,
\\
\nonumber
  \frac{ 1 }{ \sqrt{3} } \,
  \big\{
     \lvert t_0, \, e \rangle_{\mathrm{cb}}
   + \lvert t_1, \, e \rangle_{\mathrm{cb}}
&  + \lvert t_2, \, e \rangle_{\mathrm{cb}}
  \big\} ,
\quad
  \frac{ 1 }{ \sqrt{3} } \,
  \big\{
     \lvert t_0, \, t_0 \rangle_{\mathrm{cb}}
   + \lvert t_1, \, t_1 \rangle_{\mathrm{cb}}
   + \lvert t_2, \, t_2 \rangle_{\mathrm{cb}}
  \big\} ,
\\
  \frac{ 1 }{ \sqrt{2} } \,
  \big\{
     \lvert c_+, \, e \rangle_{\mathrm{cb}}
   + \lvert c_-, \, e \rangle_{\mathrm{cb}}
  \big\} ,
\quad
& \frac{ 1 }{ \sqrt{2} } \,
  \big\{
     \lvert c_+, \, c_+ \rangle_{\mathrm{cb}}
   + \lvert c_-, \, c_- \rangle_{\mathrm{cb}}
  \big\} ,
\quad
  \frac{ 1 }{ \sqrt{2} } \,
  \big\{
     \lvert c_+, \, c_- \rangle_{\mathrm{cb}}
   + \lvert c_-, \, c_+ \rangle_{\mathrm{cb}}
  \big\}
\; .
\end{align}
As expected, there are as many as there are topological charge labels.

We can rearrange the vacuum states
(\ref{eq:groundstatessmalltorussthree}) into the $\mathrm{h}$-basis
labelled by topological charges associated with the horizontal loop in
the following way:
\begin{align}\label{eq:hbasisgroundstatessmalltorussthree}
\nonumber
  \lvert (e \pm) \rangle_{\mathrm{h}}
&=
  \frac{ 1 }{ \sqrt{6} } \,
  \big\{
       \lvert e, \, e   \rangle_{\mathrm{cb}}
   \pm \lvert e, \, t_0 \rangle_{\mathrm{cb}}
   \pm \lvert e, \, t_1 \rangle_{\mathrm{cb}}
   \pm \lvert e, \, t_2 \rangle_{\mathrm{cb}}
   +   \lvert e, \, c_+ \rangle_{\mathrm{cb}}
   +   \lvert e, \, c_- \rangle_{\mathrm{cb}}
  \big\} ,
\\
\nonumber
  \lvert (e 2) \rangle_{\mathrm{h}}
&=
  \frac{ 1 }{ \sqrt{6} } \,
  \big\{
   2 \lvert e, \, e   \rangle_{\mathrm{cb}}
   - \lvert e, \, c_+ \rangle_{\mathrm{cb}}
   - \lvert e, \, c_- \rangle_{\mathrm{cb}}
  \big\} ,
\\
\nonumber
  \lvert (t \pm) \rangle_{\mathrm{h}}
&=
  \frac{ 1 }{ \sqrt{6} } \,
  \big\{
       \lvert t_0, \, e   \rangle_{\mathrm{cb}}
   \pm \lvert t_0, \, t_0 \rangle_{\mathrm{cb}}
   +   \lvert t_1, \, e   \rangle_{\mathrm{cb}}
   \pm \lvert t_1, \, t_1 \rangle_{\mathrm{cb}}
   +   \lvert t_2, \, e   \rangle_{\mathrm{cb}}
   \pm \lvert t_2, \, t_2 \rangle_{\mathrm{cb}}
  \big\} ,
\\
  \lvert (c \eta) \rangle_{\mathrm{h}}
&=
  \frac{ 1 }{ \sqrt{6} } \,
  \big\{
              \lvert c_+, \, e   \rangle_{\mathrm{cb}}
   + \bar\eta \lvert c_+, \, c_+ \rangle_{\mathrm{cb}}
   + \eta     \lvert c_+, \, c_- \rangle_{\mathrm{cb}}
   +          \lvert c_-, \, e   \rangle_{\mathrm{cb}}
   + \eta     \lvert c_-, \, c_+ \rangle_{\mathrm{cb}}
   + \bar\eta \lvert c_-, \, c_- \rangle_{\mathrm{cb}}
  \big\}
\; .
\end{align}
These are exactly the states (\ref{eq:basisvacuumhorizontal}).  The
construction of the $\mathrm{v}$-basis (\ref{eq:basisvacuumvertical})
is analogous, and the overlaps of the elements of both bases
determines the $S$-matrix of the $\mathrm{D} ( \mathrm{S}_3 )$
model~\cite{Overbosch}, which we list in table
\ref{table:smatrixdsthree}.
\begin{table}[htbp]
\centering
\begin{tabular}{c||c|c|c|c|c|c|c|c|}
  $6 \times S_{ ( C', \, \alpha' ), \, (C, \, \alpha) }$
  & \rotatebox{90}{
      $\lvert (e +) \rangle_{ \mathrm{h} }$}
  & \rotatebox{90}{
      $\lvert (e -) \rangle_{ \mathrm{h} }$}
  & \rotatebox{90}{
      $\lvert (e 2) \rangle_{ \mathrm{h} }$}
  & \rotatebox{90}{
      $\lvert (t +) \rangle_{ \mathrm{h} }$}
  & \rotatebox{90}{
      $\lvert (t -) \rangle_{ \mathrm{h} }$}
  & \rotatebox{90}{
      $\lvert (c 1)          \rangle_{ \mathrm{h} }$}
  & \rotatebox{90}{
      $\lvert (c \omega)     \rangle_{ \mathrm{h} }$}
  & \rotatebox{90}{
      $\lvert (c \bar\omega) \rangle_{ \mathrm{h} }$}
\\
\hline\hline
  ${}_{ \mathrm{v} } \langle (e +) \rvert$
  & $1$ & $1$ & $2$ & $3$ & $3$ & $2$ & $2$ & $2$ 
\\
\hline
  ${}_{ \mathrm{v} } \langle (e -) \rvert$
  & $1$ & $1$ & $2$ & $-3$ & $-3$ & $2$ & $2$ & $2$ 
\\
\hline
  ${}_{ \mathrm{v} } \langle (e 2) \rvert$
  & $2$ & $2$ & $4$ & $0$ & $0$ & $-2$ & $-2$ & $-2$ 
\\
\hline
  ${}_{ \mathrm{v} } \langle (t +) \rvert$
  & $3$ & $-3$ & $0$ & $3$ & $-3$ & $0$ & $0$ & $0$ 
\\
\hline
  ${}_{ \mathrm{v} } \langle (t -) \rvert$
  & $3$ & $-3$ & $0$ & $-3$ & $3$ & $0$ & $0$ & $0$ 
\\
\hline
  ${}_{ \mathrm{v} } \langle (c 1)          \rvert$
  & $2$ & $2$ & $-2$ & $0$ & $0$ & $4$ & $-2$ & $-2$ 
\\
\hline
  ${}_{ \mathrm{v} } \langle (c \omega)     \rvert$
  & $2$ & $2$ & $-2$ & $0$ & $0$ & $-2$ & $-2$ & $4$ 
\\
\hline
  ${}_{ \mathrm{v} } \langle (c \bar\omega) \rvert$
  & $2$ & $2$ & $-2$ & $0$ & $0$ & $-2$ & $4$ & $-2$ 
\\
\hline
\end{tabular}
\caption{\label{table:smatrixdsthree} Topological $S$-matrix for the
  $\mathrm{D} ( \mathrm{S}_3 )$ model, computed from the overlap of
  the $\mathrm{h}$-basis and the $\mathrm{v}$-basis.}
\end{table}

The $36$-dimensional two-qudit Hilbert space contains thus eight
ground states and $28$ excited states.  The latter can be
characterised as breaking the plaquette condition, the vertex
condition, or both, yielding a magnetic charge, an electric charge, or
a dyon at the single \textsl{site} defined by $P$ and $V$.  These
states we characterise as the $+1$ eigenvectors of the projectors $F_{
  \mathrm{bulk} } ( \mathcal{Q}^{ ( C, \, \alpha ) } )$.  Let us study
the excited states according to their bulk charge:
\begin{itemize}
\item %
  Bulk charge $(e -)$, that is, pure electric charges of type $1_-$:
  four states with horizontal charge labels $(e 2)$, and all of the
  $(c \eta)$:
\begin{align}\label{eq:smalltorussthreebulkoneminus}
\nonumber
& \lvert (e -)_{ \mathrm{bulk} }; \, 
         (e 2)_{ \mathrm{hor} } \rangle
=
  \frac{ 1 }{ \sqrt{2} } \,
  \big\{
     \lvert e, \, c_+ \rangle_{\mathrm{cb}}
   - \lvert e, \, c_- \rangle_{\mathrm{cb}}
  \big\} ,
\\
\nonumber
& \lvert (e -)_{ \mathrm{bulk} }; \, 
         (c \eta)_{ \mathrm{hor} } \rangle
\\
&\quad
=
  \frac{ 1 }{ \sqrt{6} } \,
  \big\{
              \lvert c_+, \, e   \rangle_{\mathrm{cb}}
   -          \lvert c_-, \, e   \rangle_{\mathrm{cb}}
   + \bar\eta \lvert c_+, \, c_+ \rangle_{\mathrm{cb}}
   - \bar\eta \lvert c_-, \, c_- \rangle_{\mathrm{cb}}
   + \eta     \lvert c_+, \, c_- \rangle_{\mathrm{cb}}
   - \eta     \lvert c_-, \, c_+ \rangle_{\mathrm{cb}}
  \big\}
\; .
\end{align}
  Notice how the horizontal charges are those which can remain
  unchanged upon fusion with the bulk charge.
\item %
  Charge $(e 2)$ (pure electric charge of type $2$): six states
  organised into three two-dimensional irreducible spaces:
\begin{align}\label{eq:smalltorussthreebulktwo}
\nonumber
& \lvert (e 2: \, v_1 )_{ \mathrm{bulk} }; \, 
         (e 2)_{ \mathrm{hor} } \rangle
=
  \frac{ 1 }{ \sqrt{3} } \,
  \big\{
                \lvert e, \, t_0 \rangle_{\mathrm{cb}}
   + \omega     \lvert e, \, t_1 \rangle_{\mathrm{cb}}
   + \bar\omega \lvert e, \, t_2 \rangle_{\mathrm{cb}}
  \big\} ,
\\
\nonumber
& \lvert (e 2: \, v_2 )_{ \mathrm{bulk} }; \, 
         (e 2)_{ \mathrm{hor} } \rangle
=
  \frac{ 1 }{ \sqrt{3} } \,
  \big\{
                \lvert e, \, t_0 \rangle_{\mathrm{cb}}
   + \bar\omega \lvert e, \, t_1 \rangle_{\mathrm{cb}}
   + \omega     \lvert e, \, t_2 \rangle_{\mathrm{cb}}
  \big\} ,
\\
\nonumber
& \lvert (e 2: \, v_1 )_{ \mathrm{bulk} }; \, 
         (t \pm)_{ \mathrm{hor} } \rangle
\\
\nonumber
&\quad
=
  \frac{ 1 }{ \sqrt{6} } \,
  \big\{
                  \lvert t_0, \, e \rangle_{\mathrm{cb}}
   +   \omega     \lvert t_1, \, e \rangle_{\mathrm{cb}}
   +   \bar\omega \lvert t_2, \, e \rangle_{\mathrm{cb}}
   \pm            \lvert t_0, \, t_0 \rangle_{\mathrm{cb}}
   \pm \omega     \lvert t_1, \, t_1 \rangle_{\mathrm{cb}}
   \pm \bar\omega \lvert t_2, \, t_2 \rangle_{\mathrm{cb}}
  \big\} ,
\\
\nonumber
& \lvert (e 2: \, v_2 )_{ \mathrm{bulk} }; \, 
         (t \pm)_{ \mathrm{hor} } \rangle
\\
&\quad
=
  \frac{ 1 }{ \sqrt{6} } \,
  \big\{
                  \lvert t_0, \, e \rangle_{\mathrm{cb}}
   +   \bar\omega \lvert t_1, \, e \rangle_{\mathrm{cb}}
   +   \omega     \lvert t_2, \, e \rangle_{\mathrm{cb}}
   \pm            \lvert t_0, \, t_0 \rangle_{\mathrm{cb}}
   \pm \bar\omega \lvert t_1, \, t_1 \rangle_{\mathrm{cb}}
   \pm \omega     \lvert t_2, \, t_2 \rangle_{\mathrm{cb}}
  \big\}
\; .
\end{align}
  Again the only horizontal labels are the irreps that can fuse with
  $(e 2)$ without change, namely $(e 2)$ and $(t \pm)$.
\item %
  The charges $(t \pm)$ never appear in the bulk.  The transpositions
  $t_0$, $t_1$, $t_2$ do not belong to the commutator group of
  $\mathrm{S}_3$; on the other hand, these representations always
  change the charges that they fuse with.
\item %
  Each charge $(c 1)$, $(c \omega)$, $(c \bar\omega)$, whose magnetic
  part is a flux of $3$-cycle type, appears in the bulk in six states,
  organised into three two-dimensional modules.  Let $\eta$, as usual,
  run over the cubic roots of unity $1, \, \omega, \, \bar\omega$.
  Then
\begin{align}\label{eq:smalltorussthreebulkconeplus}
\nonumber
& \lvert (c \eta: \, c_+ )_{ \mathrm{bulk} }; \, 
         (t \pm)_{ \mathrm{hor} } \rangle
\\
\nonumber
&\quad
=
  \frac{ 1 }{ \sqrt{6} } \,
  \big\{
                \lvert t_0, \, t_1 \rangle_{\mathrm{cb}}
   +   \bar\eta \lvert t_1, \, t_2 \rangle_{\mathrm{cb}}
   +   \eta     \lvert t_2, \, t_0 \rangle_{\mathrm{cb}}
   \pm          \lvert t_0, \, c_+ \rangle_{\mathrm{cb}}
   \pm \bar\eta \lvert t_1, \, c_+ \rangle_{\mathrm{cb}}
   \pm \eta     \lvert t_2, \, c_+ \rangle_{\mathrm{cb}}
  \big\} ,
\\
\nonumber
& \lvert (c \eta: \, c_- )_{ \mathrm{bulk} }; \, 
         (t \pm)_{ \mathrm{hor} } \rangle
\\
\nonumber
&\quad
=
  \frac{ 1 }{ \sqrt{6} } \,
  \big\{
                \lvert t_0, \, t_2 \rangle_{\mathrm{cb}}
   +   \eta     \lvert t_1, \, t_0 \rangle_{\mathrm{cb}}
   +   \bar\eta \lvert t_2, \, t_1 \rangle_{\mathrm{cb}}
   \pm          \lvert t_0, \, c_- \rangle_{\mathrm{cb}}
   \pm \eta     \lvert t_1, \, c_- \rangle_{\mathrm{cb}}
   \pm \bar\eta \lvert t_2, \, c_- \rangle_{\mathrm{cb}}
  \big\} ,
\\
\nonumber
& \lvert (c \eta: \, c_+ )_{ \mathrm{bulk} }; \, 
         (c \eta)_{ \mathrm{hor} } \rangle
=
  \frac{ 1 }{ \sqrt{3} } \,
  \big\{
              \lvert c_-, \, t_0 \rangle_{\mathrm{cb}}
   + \bar\eta \lvert c_-, \, t_1 \rangle_{\mathrm{cb}}
   + \eta     \lvert c_-, \, t_2 \rangle_{\mathrm{cb}}
  \big\} ,
\\
& \lvert (c \eta: \, c_- )_{ \mathrm{bulk} }; \, 
         (c \eta)_{ \mathrm{hor} } \rangle
=
  \frac{ 1 }{ \sqrt{3} } \,
  \big\{
              \lvert c_+, \, t_0 \rangle_{\mathrm{cb}}
   + \eta     \lvert c_+, \, t_1 \rangle_{\mathrm{cb}}
   + \bar\eta \lvert c_+, \, t_2 \rangle_{\mathrm{cb}}
  \big\}
\; .
\end{align}
Indeed, the only charges that can fuse with $(c \eta)$ without change
are $(c \eta)$ itself and the $(t \pm)$.
\end{itemize}

This label distribution is summarised in table
\ref{table:bulkhorchargesdsthree}.
\begin{table}[htbp]
\centering
\begin{tabular}{c||c|c|c|c|c|c|c|c||c|c}
    bulk $\backslash$ loop
  & $(e +)$ 
  & $(e -)$ 
  & $(e 2)$ 
  & $(t +)$ 
  & $(t -)$ 
  & $(c 1         )$ 
  & $(c \omega    )$ 
  & $(c \bar\omega)$ 
  & total & energy
\\
\hline\hline
  $(e +)$
  & $1$ & $1$ & $1$ & $1$ & $1$ & $1$ & $1$ & $1$ 
  & $8$ & $-2$
\\
\hline
  $(e -)$
  & $-$ & $-$ & $1$ & $-$ & $-$ & $1$ & $1$ & $1$ 
  & $4$ & $-1$
\\
\hline
  $(e 2)$
  & $-$ & $-$ & $2$ & $2$ & $2$ & $-$ & $-$ & $-$ 
  & $6$ & $-1$
\\
\hline
  $(t +)$
  & $-$ & $-$ & $-$ & $-$ & $-$ & $-$ & $-$ & $-$ 
  & $0$ & $(-1)$
\\
\hline
  $(t -)$
  & $-$ & $-$ & $-$ & $-$ & $-$ & $-$ & $-$ & $-$ 
  & $0$ & $(0)$
\\
\hline
  $(c 1)$
  & $-$ & $-$ & $-$ & $2$ & $2$ & $2$ & $-$ & $-$ 
  & $6$ & $-1$
\\
\hline
  $(c \omega)$
  & $-$ & $-$ & $-$ & $2$ & $2$ & $-$ & $2$ & $-$ 
  & $6$ & $0$
\\
\hline
  $(c \bar\omega)$
  & $-$ & $-$ & $-$ & $2$ & $2$ & $-$ & $-$ & $2$ 
  & $6$ & $0$
\\
\hline
\end{tabular}
\caption{\label{table:bulkhorchargesdsthree} Bulk and horizontal loop
  charge label pairs realised on the small torus.  The entries
  represent the number of states belonging to the corresponding charge
  pair.}
\end{table}

For each bulk charge, an analogue of the topological $S$-matrix can be
constructed by computing the overlaps of elements of the horizontal
and vertical bases: the results are gathered in table
\ref{table:smatrixexcited}.
\begin{table}[htbp]
\centering
\begin{tabular}{c||c|c|c|c|}
  $\sqrt{3} \times S [ (e -)_{ \mathrm{bulk} } ]$
  & $\lvert (e 2) \rangle_{ \mathrm{h} }$
  & $\lvert (c 1          ) \rangle_{ \mathrm{h} }$
  & $\lvert (c \omega     ) \rangle_{ \mathrm{h} }$
  & $\lvert (c \bar\omega ) \rangle_{ \mathrm{h} }$
\\
\hline\hline
  ${}_{ \mathrm{v} } \langle (e 2) \rvert$
  & $0$ & $1$ & $1$ & $1$
\\
\hline
  ${}_{ \mathrm{v} } \langle (c 1) \rvert$
  & $1$ & $0$ & $-i$ & $i$
\\
\hline
  ${}_{ \mathrm{v} } \langle (c \omega) \rvert$
  & $1$ & $-i$ & $i$ & $0$
\\
\hline
  ${}_{ \mathrm{v} } \langle (c \bar\omega) \rvert$
  & $1$ & $i$ & $0$ & $-i$
\\
\hline
\end{tabular}

\vspace{1cm}

\begin{tabular}{c||c|c|c|}
  $2 \times S [ (e 2)_{ \mathrm{bulk} } ]$
  & $\lvert (e 2) \rangle_{ \mathrm{h} }$
  & $\lvert (t +) \rangle_{ \mathrm{h} }$
  & $\lvert (t -) \rangle_{ \mathrm{h} }$
\\
\hline\hline
  ${}_{ \mathrm{v} } \langle (e 2) \rvert$
  & $0$ & $\sqrt{2}$ & $\sqrt{2}$
\\
\hline
  ${}_{ \mathrm{v} } \langle (t +) \rvert$
  & $\sqrt{2}$ & $1$ & $-1$
\\
\hline
  ${}_{ \mathrm{v} } \langle (t -) \rvert$
  & $\sqrt{2}$ & $-1$ & $1$
\\
\hline
\end{tabular}

\vspace{1cm}

\begin{tabular}{c||c|c|c|}
  $2 \times S [ (c \eta)_{ \mathrm{bulk} } ]$
  & $\lvert (c \eta) \rangle_{ \mathrm{h} }$
  & $\lvert (t +) \rangle_{ \mathrm{h} }$
  & $\lvert (t -) \rangle_{ \mathrm{h} }$
\\
\hline\hline
  ${}_{ \mathrm{v} } \langle (c \eta) \rvert$
  & $0$ & $\sqrt{2}$ & $- \sqrt{2}$
\\
\hline
  ${}_{ \mathrm{v} } \langle (t +) \rvert$
  & $\sqrt{2} \, \bar\eta $ & $1$ & $1$
\\
\hline
  ${}_{ \mathrm{v} } \langle (t -) \rvert$
  & $- \sqrt{2} \, \bar\eta$ & $1$ & $1$
\\
\hline
\end{tabular}

\vspace{1cm}

\caption{\label{table:smatrixexcited} Analogues of the topological
  $S$-matrix for the bulk excited states in the $\mathrm{D} (
  \mathrm{S}_3 )$ model.  By adjusting the phases in the definition of
  the bases in the last table we can obtain exactly the coefficients
  of the second table (all two-dimensional irreps of $\mathrm{D} (
  \mathrm{S}_3 )$ have the same behaviour with respect to modular
  transformations).}
\end{table}

\clearpage


\end{document}